\documentstyle[12pt,epsfig]{report}
\def\thebibliography#1{\leftline{\Large\it References}\list
  {[\arabic{enumi}]}{\settowidth\labelwidth{[#1]}\leftmargin\labelwidth
    \advance\leftmargin\labelsep
    \usecounter{enumi}}
    \def\newblock{\hskip .11em plus .33em minus .07em}
    \sloppy\clubpenalty4000\widowpenalty4000}

\newcommand{\be}{\begin{eqnarray}}
\newcommand{\ba}{\begin{array}}
\newcommand{\ea}{\end{array}}
\newcommand{\ee}{\end{eqnarray}}

\newcommand{\dslash}{\partial \hskip -0.5em /}
\newcommand{\kslash}{k \hskip -0.5em /}
\newcommand{\vslash}{v \hskip -0.5em /}
\newcommand{\aslash}{a \hskip -0.5em /}
\newcommand{\pslash}{p \hskip -0.5em /}
\newcommand{\qslash}{q \hskip -0.5em /}

\newcommand{\nslash}{{\rm n} \hskip -0.5em /}
\newcommand{\bD}{{\bf D}}
\newcommand{\bDp}{{\bf D}^{(\pi)}}

\newcommand{\La}{{\cal L}}
\newcommand{\A}{{\cal A}}

\newcommand{\bjlim}{{\stackrel{\scriptstyle{\rm Bj}}
{\textstyle\longrightarrow}}}

\newcommand{\ID}{\mbox{{\sf 1}\hskip-0.5mm
\rule{0.04em}{1.55ex}\hskip0.2mm}}

\newcommand{\xipl}{\vec{\xi}\hskip-0.6mm
+\hskip-0.6mm\lambda\hat{e}_3}
\newcommand{\ximl}{\vec{\xi}\hskip-0.6mm
-\hskip-0.6mm\lambda\hat{e}_3}
\newcommand{\tauom}{\vec{\tau}\hskip-0.3mm
\cdot\hskip-0.3mm\vec{\Omega}}

\begin{document}
%\vskip -2.0truecm
%\centerline{\fbox{\parbox[t]{3.0cm}
%{\begin{center}
%{\large \bf Draft!}\\
%regbf.tex\\ \today
%\end{center}}}}
\vskip -1.0truecm
\rightline{\footnotesize{MIT--CTP--2861}}
\rightline{\footnotesize{OKHEP--99--03}}
\rightline{\footnotesize{hep-ph/9905329}}
\vskip 0.2truecm
\centerline{\Large\bf Hadron Structure Functions in a Chiral Quark Model:}
\vskip0.3cm
\centerline{\Large\bf
Regularization, Scaling and Sum Rules$^{\textstyle*}$}
\baselineskip=14 true pt
\vskip 0.5cm
\centerline{H.\ Weigel$^{{\textstyle\dagger}\,{\rm a}}$,
E. Ruiz Arriola$^{\rm b}$ and L. Gamberg$^{\rm c}$}
\vskip 0.3cm
\begin{center}
$^{\rm a}$Center for Theoretical Physics\\
Laboratory of Nuclear Science and Department of Physics\\
Massachusetts Institute of Technology\\
Cambridge, Ma 02139
\vskip 0.2cm
$^{\rm b}$Departamento de F\'{\i}sica Moderna\\
Universidad de Granada\\
E--18071 Granada, Spain
\vskip 0.2cm
$^{\rm c}$Department of Physics and Astronomy\\
University of Oklahoma\\
440 West Brooks, Norman, Ok 73019
\end{center}
\vskip 0.4cm
\baselineskip=14pt
\centerline{\bf ABSTRACT}
\vskip 0.3cm
\centerline{\parbox[t]{15.8cm}{\baselineskip=14pt
We provide a consistent regularization procedure for calculating 
hadron structure functions in a chiral quark model.
The structure functions are extracted from the absorptive part 
of the forward Compton amplitude in the Bjorken limit.
Since this amplitude is obtained as a time--ordered correlation
function its regularization is consistently determined from the 
regularization of the bosonized action. We find that the Pauli--Villars
regularization scheme is most suitable because it preserves both
the anomaly structure of QCD and the leading scaling behavior of 
hadron structure functions in the Bjorken limit. We show that this
procedure yields the correct pion structure function. In order to 
render the sum rules of the regularized polarized nucleon structure 
functions consistent with their corresponding axial charges we find 
it mandatory to further specify the regularization procedure. 
This specification goes beyond the double subtraction scheme commonly
employed when studying static hadron properties in this model. 
In particular the present approach serves to determine the 
regularization prescription for structure functions whose leading 
moments are not given by matrix elements of local operators. In this 
regard we conclude somewhat surprisingly that in this model 
the Gottfried sum rule does not undergo regularization.}}
\vskip 0.15cm
\leftline{{\footnotesize{\it PACS: 11.30.Rd, 11.55.He, 
12.39.-x, 12.39.Fe}}}
\leftline{\footnotesize{\it Keywords: 
\parbox[t]{13.5cm}{Structure functions,
Compton amplitude, chiral symmetry, Pauli--Villars regularization,
soliton model, collective quantization, sum rules}}}

\vfill
\noindent
--------------

\noindent
$^{\textstyle*}$\hskip-0.03cm
{\footnotesize This work is supported in parts by funds provided by the 
U.S. Department of Energy (D.O.E.) under cooperative research
agreements \#DF--FC02--94ER40818, and \#DE--FG03--98ER41066, 
the Deutsche Forschungsgemeinschaft (DFG) 
under contract We 1254/3-1 and the Spanish DGES grant no. PB95-1204 
and the Junta de Andaluc{\'{\i}}a grant no. FQM0225}.\\
$^{\textstyle\dagger}$\hskip-0.02cm
{\footnotesize {Heisenberg--Fellow}}
\eject

\baselineskip=16pt

\stepcounter{chapter}
\leftline{\Large\it 1. Introduction}
\medskip

The analysis of deep inelastic scattering (DIS) provides some 
of the most convincing evidence for the quark sub--structure 
\cite{Ro90} of the nucleon because the resulting cross--sections
are parameterized by structure functions which exhibit scaling 
properties characteristic of the quark sub--structure. 
The parton model parameterizes these structure functions in 
terms of quark and anti--quark distributions. To leading order 
in an expansion in twist perturbative QCD in conjunction with 
the quark--parton model successfully describes the observed violations 
of the scaling properties. In addition, using the world data 
the various parameterizations \cite{Gl95} of quark distributions 
are successfully categorizing a wealth of information about the 
(quark) sub--structure of hadrons. 

However, without bound state wave--functions from first principles in QCD 
for  hadrons it is not possible to calculate the quark distributions
and corresponding hadron structure functions. On the other hand,
from the point of view of an effective field theory the study of 
quark momentum distributions in a hadron is particularly challenging, 
because it requires the coexistence of hadronic and quark degrees of 
freedom, without perpetration of double counting.  As of present
there is no obvious expansion parameter which facilitates a 
rigorous accounting of such contributions.  This is to be contrasted 
with the very low or very high energy regimes where respectively the 
chiral expansion or perturbative QCD apply.  Thus far, one must resort 
to model calculations \cite{Ja75,model}.  In this context it is important 
to  note that there are the phenomenologically successful chiral models 
for hadron properties \cite{soliton}.  From the perspective that hadron 
structure functions can be viewed as functions of quark distributions we 
adopt the Nambu--Jona--Lasinio (NJL) model \cite{Na61} of quark flavor 
dynamics. This model offers a microscopic description of dynamical 
chiral symmetry breaking for low energy non--perturbative hadron  physics.   
Upon bosonization of the quark fields, meson fields are described as 
functionals in terms of non--perturbative quark anti--quark bound 
states \cite{Eb86,Vo91}.  Moreover chiral soliton configurations can be 
constructed for these meson fields \cite{Al96}.  In this manner the model 
provides a non--perturbative description not only for the mesons but 
also the baryons. In the case of the latter the quark fields become 
functionals of a chiral soliton. During the process of bosonization a 
link to the quark degrees of freedom is maintained and one is tempted 
to compare the associated quark distributions to those which 
enter the quark--parton model description of structure functions. 
The major reason, however, for adopting this model is that the hadronic 
tensor, which defines the structure functions, can efficiently
be studied in this model. This is based on the observation 
that prior to bosonization the symmetry currents entering 
the hadronic tensor are formally identical to currents in a 
non--interacting Dirac theory. At this formal level the NJL 
quark distributions can indeed be identified with 
those of the parton model. However, a major obstacle is 
that the bosonized NJL model contains quadratic divergences 
which have to be regularized. This regularization eventually 
spoils the formal identity of these distributions. For that reason 
the regularization of pion structure functions has sometimes 
been imposed {\it a posteriori}.

In the NJL chiral soliton model the issue of systematically regularizing 
the nucleon structure functions has yet to be addressed. Rather 
various approximations have been made. In the first approximation the 
contributions of the polarized vacuum to the structure functions were 
omitted \cite{We96,We97}. This approach is motivated by the 
observation that (after regularization) the vacuum contributions to 
the corresponding static nucleon properties are small once the 
self--consistent soliton is constructed. Other approaches included 
vacuum contributions, however, regularization was introduced in a 
somewhat {\it ad hoc} manner by performing a single Pauli--Villars 
type subtraction \cite{Di96,Wa98,Po99}. The single subtraction 
Pauli--Villars treatment is limited to the chirally symmetric 
case of vanishing current quark masses as this particular case 
allows one to ignore the quadratically divergent gap equation 
which is not rendered finite by a single subtraction.  In addition, 
attempts to impose a regularization scheme at the level of quark 
distributions using error--function and Gaussian damping of the 
ultraviolet behavior was carried out in ref \cite{Wa98}\footnote{As 
the authors carefully noted these regularization schemes were not 
intended to be self consistent. Similar schemes were applied 
in \cite{Ta96}.}. In turn the regularization attributed 
to a given structure function was conjectured to equal 
that of the corresponding sum rule associated with
static nucleon properties. This connection to static 
properties for setting up the regularization description 
does not provide a definite answer for structure functions 
not having a sum rule; {\it i.e.}, structure functions whose
integral cannot be written as a matrix element of a local
operator. The major purpose of the present 
investigation is to establish a consistent regularization
procedure for obtaining nucleon structure functions in the 
NJL model. Also, it happens that the issue of regularization has 
significant impact for the pion structure functions, thus
we will  discuss that case as well. 

Before bosonization the NJL  model is defined in terms
of quark fields by the Lagrangian~\cite{Na61}
\be
\La_{\rm NJL} = \bar q (i\dslash - m_0 ) q +
      2G_{\rm NJL} \left\{ (\bar q \frac{\vec{\tau}}{2} q )^2
      +(\bar q \frac{\vec{\tau}}{2} i\gamma _5 q )^2 \right\}
\label{NJL}
\ee
with the quark interaction described by a chirally symmetric 
quartic potential. The current quark mass, $m_0$, parameterizes the 
small explicit breaking of chiral symmetry. Using functional 
techniques the quark fields can be integrated out in favor of 
auxiliary mesonic fields, ${\cal M}=S+iP$. According to the 
chirally symmetric interaction in the Lagrangian (\ref{NJL}),
$S$ and $P$ are scalar and pseudoscalar degrees of freedom, 
respectively. This results in the bosonized action~\cite{Eb86}
\be
\A_{\rm NJL}=-iN_C {\rm Tr}_\Lambda\, {\rm log}\,
\left\{i\dslash - m_0 - \left(S + i\gamma_5 P\right)\right\}
+\frac{1}{4G}\int d^4x\,
{\rm tr}\left[{\cal M}{\cal M}^\dagger\right]\, .
\label{act0}
\ee
Here the `cut--off' $\Lambda$ indicates that the 
quadratically divergent quark loop requires regularization.
In order to compute properties of hadrons from the 
action (\ref{act0}) a twofold procedure is in order. First, 
formal expressions for the symmetry currents have
to be extracted. This is straightforwardly accomplished
by adding external sources to the Dirac operator and
taking the appropriate functional derivatives. Ignoring effects
associated with the regularization, the currents would 
be as simple as $\bar{q}\gamma_\mu (\gamma_5) t^a q$, with 
$t^a$ being the appropriate flavor generator. Secondly,
hadron states are constructed from the effective
action which in turn allows one to calculate the relevant 
matrix elements of the symmetry currents. For
the pion this will be a Bethe--Salpeter wave--function
which is obtained by expanding the action (\ref{act0})
appropriately in the fields $S$ and $P$. In the case
of the nucleon we will determine a soliton configuration
which after collective quantization carries nucleon quantum
numbers \cite{ANW}. These issues will be reviewed in section 2.

For hadron structure functions we are required to evaluate
matrix elements of bi--local current correlation functions
which define the hadronic tenor. As is well--known, this tensor 
parameterizes the strong interaction part of a deep inelastic 
scattering (DIS) cross--section and is decomposed into form factors. 
The form factors depend only on the Lorentz invariants $Q^2=-q^2$ 
and $x=Q^2/2p\cdot q$ where $p$ is the hadron momentum and $q$ 
denotes the momentum transferred to the hadron by the exchange 
of a virtual gauge boson. In general the hadronic tensor also 
depends on the spin orientation, $s$, of the hadron as well as 
the flavor quantum numbers of the hadronic current $J_{\mu}^{a}$ 
to which the exchanged boson couples,
\be
W_{\mu \nu}^{ab}(p,q;s) = \frac{1}{4\pi}
\int d^4x e^{iq\cdot \xi}\,
\Big\langle p,s\Big| [ J_{\mu}^{a}(\xi),J_{\nu}^{b \dagger}(0)]
\Big| p,s \Big\rangle\, .
\label{hten1}
\ee
As the currents are rigorously determined within the model 
the hadronic tensor is a well--defined quantity and except 
for calculational matters no further assumptions are needed 
to compute~$W_{\mu\nu}$.

In the case of the nucleon, the form factors appear in the 
Lorentz--covariant decomposition 
(omitting parity violating contributions)
\be
W_{\mu \nu}^{ab}(p,q;s)
& = & \left(-g_{\mu \nu} + \frac{q_{\mu} q_{\nu}}{q^2}\right)
M_N W_{1} (x,Q^2)
\nonumber \\ & &
+\left(p_{\mu} - q_{\mu}\frac{p\cdot q}{q^2}\right)
\left(p_{\nu} - q_{\nu}\frac{p\cdot q}{q^2}\right)
\frac{1}{M_N} W_{2} (x,Q^2)
\nonumber \\ & &
+i\epsilon_{\mu \nu \lambda \sigma} \frac{q^{\lambda} M_N}{p\cdot q}
\left(\left[g_1(x,Q^2)+g_2(x,Q^2)\right]s^{\sigma}
-\frac{q\cdot s}{q\cdot p} p^{\sigma}
    g_2(x,Q^2) \right)\, .
\label{hten1a}
\ee
When a spin--zero hadron is considered, as {\it e.g.} the
pion, the polarized structure functions $g_1$ and $g_2$ are ignored,
{\it cf.} eq (\ref{disp1}).
Once the hadronic tensor is computed, the form factors are given 
by suitable projections. Finally the leading twist contributions
to the structure functions are obtained from these form factors by 
assuming the Bjorken limit: 
\be
Q^2\to\infty
\qquad {\rm with}\qquad
x=Q^2/2p\cdot q \quad {\rm fixed}\, .
\label{bjl}
\ee
For the spin independent part the structure functions $f_i$ are the
linear combinations
\be
M_N W_1(x,Q^2)\, \bjlim\,  f_1(x)
\quad {\rm and} \quad
\frac{p\cdot q}{M_N}\,W_2(x,Q^2)\, \bjlim\,  f_2(x)\, .
\label{deff1f2}
\ee

However, the expression (\ref{hten1}) is not conveniently
computed from the bosonized action (\ref{act0}). In case 
the hadron represents the ground state with
the specific quantum numbers, the hadronic tensor will be
related to the forward virtual Compton amplitude
\be
W_{\mu \nu}^{ab}(p,q;s)=\frac{1}{2\pi}\, {\rm Im}\, 
T_{\mu \nu}^{ab}(p,q;s)\, ,
\label{Comp1}
\ee
which is given as the matrix element of a time--ordered
product of the currents
\be
T_{\mu \nu}^{ab}(p,q;s) = i
\int d^4\xi e^{iq\cdot \xi}\,
\Big\langle p,s\Big| T\left\{J_{\mu}^{a}(\xi) 
J_{\nu}^{b \dagger}(0)\right\}\Big| p,s \Big\rangle
\label{Comp2}
\ee
rather than their commutator. This time--ordered product can 
easily be obtained by expanding the bosonized action (\ref{act0})
up to quadratic order in the external fields. Applying Cutkosky's
rules subsequently yields the absorptive part,
${\rm Im}\, T_{\mu \nu}$, which is to be 
identified with the hadronic tensor (\ref{Comp1}).

The paper is organized as follows. In section~2 we will review
the model with regard to the meson and soliton sectors. In 
section~3 we will discuss in detail the calculation of the pion 
structure function.  In the fully regularized model this 
goes beyond simply computing the standard `handbag' diagram.
Fortunately the Bjorken limit provides some simplifications.
In section~4 will will discuss how these simplifications can
be transferred to the soliton sector; thereby exhibiting 
the leading scaling laws. Section~5 represents
the main part of the paper providing formal expressions for
the nucleon structure functions in the fully regularized 
NJL chiral soliton model. In section~6 we will perform consistency
checks with regard to the pertinent sum rules. We will summarize
our results in section~7. Starting from the Bethe--Salpeter amplitude 
an alternative computation of the pion structure function is 
presented in appendix~A while appendix~B contains a formal
discussion on the momentum distribution among the propagators 
in the quark loop. In appendix~C we will present a brief discussion 
of the (unphysical) model which has been included to regularize 
the quadratically divergent action (\ref{act0}). Finally appendix~D 
contains some lengthy expressions for the nucleon structure functions 
at sub--leading order in $1/N_C$. At this order the structure 
functions which enter the Alder and Gottfried sum rules are
non--vanishing. Surprisingly we find that the latter does not 
undergo regularization. Some preliminary results of the present 
studies have already been communicated previously \cite{Da95,We99}.

\bigskip
\stepcounter{chapter}
\leftline{\Large\it 2. Pauli--Villars regularization 
for the NJL model}
\medskip

Upon employing the Pauli--Villars regularization scheme it is 
possible to formulate the bosonized NJL model \cite{Na61} 
completely in Minkowski space. This will be quite appropriate 
when applying Cutkosky's rules in order to extract the hadronic 
tensor from the Compton amplitude. Also, let us remind the reader 
that in ref \cite{Da95}, scaling for the pion structure functions 
was accomplished in the Pauli-Villars regularization, and not in 
the proper--time regularization. In this context, it has been 
shown \cite{Da95bis} that in the Pauli-Villars regularization, unlike 
the more customary proper--time scheme where cuts in the complex
plane appear \cite{Br96,Al96}, dispersion relations are fulfilled. 

\bigskip
\leftline{\large\it 2a. Vacuum and meson sectors}
\medskip

In this subsection we will set the stage for our investigation 
of structure functions by first reviewing the vacuum and meson 
sectors of the NJL model. We will be as brief as possible because 
these issues have been thoroughly discussed in the literature 
\cite{Eb86,Vo91,Al96}. The Pauli--Villars regularization has been 
considered before in this context both for mesons and solitons 
and we refer to refs \cite{Da95,Da95bis,RA95,Ru91,Sc92,Do93} for more 
details and results in this regularization scheme. We will 
specifically follow ref \cite{Da95} because in that formulation 
a consistent treatment solely in Minkowski space is possible. 

We consider two different Dirac operators in the background of 
scalar ($S$) and pseudoscalar ($P$) fields \cite{Da95,RA95}
\be
i \bD &=& i\dslash - \left(S+i\gamma_5P\right)
+\vslash +\aslash\gamma_5
=:i\bDp+\vslash+\aslash\gamma_5
\label{defd} \\
i \bD_5 &=& - i\dslash - \left(S-i\gamma_5P\right)
-\vslash+\aslash\gamma_5
=:i\bDp_5-\vslash+\aslash\gamma_5\, .
\label{defd5}
\ee
Here we have also introduced external vector ($v_\mu$) and 
axial--vector ($a_\mu$) fields. As noted above the functional 
derivate of the action with respect to these sources will provide 
the vector and axial--vector currents, respectively. For later use 
we have also defined Dirac operators, $\bDp$ and $\bDp_5$, with 
these fields omitted. Of course, all fields appearing in eqs 
(\ref{defd}) and (\ref{defd5}) are considered to be matrix fields 
in flavor space. It is worth noting that upon continuation to 
Euclidean space, $\bD_5$ transforms into the hermitian
conjugate of $\bD$ \cite{RA95}.

The regularized action of the bosonized NJL model is then given
by\footnote{Here and henceforth we denote traces of discrete 
indices by ``${\rm tr}$'' while ``${\rm Tr}$'' also contains 
the space--time integration.}
\be
\A_{\rm NJL}&=&\A_{\rm R}+\A_{\rm I}
+\frac{1}{4G}\int d^4x\, 
{\rm tr}\left[S^2+P^2+2m_0S\right]
\label{act1} \\
\A_{\rm R}&=&-i\frac{N_C}{2}
\sum_{i=0}^2 c_i {\rm Tr}\, {\rm log}
\left[- \bD \bD_5 +\Lambda_i^2-i\epsilon\right]\, ,
\label{act2} \\
\A_{\rm I}&=&-i\frac{N_C}{2}
{\rm Tr}\, {\rm log} 
\left[-\bD \left(\bD_5\right)^{-1}-i\epsilon\right]\, .
\label{act3}
\ee
The local term in eq (\ref{act1}) is the remainder of the
quartic quark interaction of the NJL model. After having shifted
the meson fields by an amount proportional to the current 
quark masses $m_0$ it also contains the explicit breaking of 
chiral symmetry. Furthermore we have retained the notion
of real and imaginary parts of the action as it would come
about in the Euclidean space formulation. This is also indicated
by the Feynman boundary conditions. When disentangling these 
pieces, it is found that only the `real part' $\A_{\rm R}$ is 
ultraviolet divergent. It is regularized within the Pauli--Villars 
scheme according to which the conditions\footnote{In the case 
of two subtractions we need at least two
cut-offs $\Lambda_1 $ and $\Lambda_2$. In the limit 
$\Lambda_1  \to \Lambda_2 = \Lambda $, we have 
$\sum_i c_i f(\Lambda_i^2)= f(0)-f(\Lambda^2)+\Lambda^2 f' (\Lambda^2 )$.
For instance, $\sum_i c_i \Lambda_i^{2n}=(2n-2)\Lambda^{2n}$.}
\be
c_0=1\, ,\quad \Lambda_0=0\, ,\quad \sum_{i=0}^2c_i=0
\quad {\rm and}\quad \sum_{i=0}^2c_i\Lambda_i^2=0
\label{pvcond}
\ee
hold. The `imaginary part' $\A_{\rm I}$ is conditionally convergent, 
{\it i.e.} a principle value description must be imposed for the 
integration over the time coordinate. {\it A priori} this does not 
imply that it should not be regularized. However, in order to 
correctly reproduce the axial anomaly we are constrained to leave 
it unregularized \cite{Dh85,Eb86}. In the case of vector 
interaction the situation is a bit more involved because
suitable counterterms have to be added \cite{RA95,Wa96}
.

Essentially we have added and subtracted the 
(unphysical) $\bD_5$ model to the bosonized NJL model. Under 
regularization the sum, ${\rm log}\,(\bD)+{\rm log}\,(\bD_5)$ 
is then treated differently from the difference,
${\rm log}\,(\bD)-{\rm log}\,(\bD_5)$.
In the case of the polarized nucleon structure 
functions we will later recognize that this special choice 
of regularization nevertheless requires further specification.

In the vacuum sector the pseudoscalar fields vanish
while the variation of the action with respect to the 
scalar field $S$ and yields the gap equation
\be
\frac{1}{2G}\left(m-m_0\right)
=-4iN_C m\sum_{i=0}^2c_i
\int\frac{d^4k}{(2\pi)^4}
\left[-k^2+m^2+\Lambda_i^2-i\epsilon\right]^{-1}\, .
\label{gap}
\ee
This equation determines the vacuum expectation value of 
the scalar field $\langle S \rangle =m$ which is referred to
as the constituent quark mass. Its non--vanishing value
signals the dynamical breaking of chiral symmetry.
Next we expand the action to quadratic order
in the pion field ${\vec\pi}$. This field resides in the 
non--linear representation of the meson fields on the 
chiral circle
\be
{\cal M}=m\, U = m\, {\rm exp}
\left(i \frac{g}{m}\,{\vec\pi} \cdot {\vec\tau} \right)\, .
\label{chifield}
\ee
This representation also defines the chiral field $U$. The 
quark--pion coupling $g$ will be specified shortly. Upon Fourier 
transforming to ${\vec{\tilde\pi}}$ we find
\be
\A_{\rm NJL}=g^2\int \frac{d^4q}{(2\pi)^4}\,
{\vec{\tilde\pi}}(q) \cdot {\vec{\tilde\pi}}(-q)
\left[2N_C q^2\Pi(q^2)-\frac{1}{2G}\frac{m_0}{m}\right]
+{\cal O}\left(\vec{\pi}^4\right)\, ,
\label{A2}
\ee
with the polarization function
\be
\Pi(q^2,x)&=&-i\sum_{i=0}^2 c_i\,
\frac{d^4k}{(2\pi)^4}\,
\left[-k^2-x(1-x)q^2+m^2+\Lambda_i^2-i\epsilon\right]^{-2}
\quad {\rm and} \quad
\nonumber \\
\Pi(q^2)&=&\int_0^1 dx\, \Pi(q^2,x)\, ,
\label{specfct}
\ee
parameterizing the quark loop. The on--shell condition
for the pion relates its mass to the model 
parameters
\be
m_\pi^2=\frac{1}{2G}\frac{m_0}{m}\frac{1}{2N_C\Pi(m_\pi^2)}\, .
\label{mpi}
\ee
Requiring a unit residuum at the pion pole
determines the quark--pion coupling
\be
\frac{1}{g^2}=4N_C \frac{d}{dq^2}
\left[q^2 \Pi(q^2)\right]\Bigg|_{q^2=m_\pi^2} \, .
\label{gpcoup}
\ee
The axial current is obtained from the linear coupling to
the axial--vector source $a_\mu$. Its matrix element between
the vacuum and the properly normalized one--pion state 
provides the pion decay constant $f_\pi$ as a function of the 
model parameters
\be
f_\pi=4N_C m g \Pi(m_\pi^2)\, .
\label{fpi}
\ee
The empirical values $m_\pi=138{\rm MeV}$ and $f_\pi=93{\rm MeV}$ 
are used to determine the model parameters.

\bigskip
\leftline{\large\it 2b. Soliton sector}
\medskip
In order to describe a soliton configuration we consider
static meson configurations. In that case it is suitable
to introduce a Dirac Hamiltonian $h$ via
\be
i\bDp=\beta(i\partial_t-h) \quad {\rm and}\quad
i\bDp_5=(-i\partial_t-h)\beta\, .
\label{defh}
\ee
For a given meson configuration the Hamiltonian $h$ is 
diagonalized
\be
h\Psi_\alpha = \epsilon_\alpha \Psi_\alpha \, ,
\label{diagh}
\ee
yielding eigen--spinors $\Psi_\alpha$ and energy eigenvalues
$\epsilon_\alpha$. In the unit baryon number sector the well--known 
hedgehog configuration minimizes the action for the meson fields. 
This configuration introduces the chiral angle $\Theta(r)$ via
\be
h=\vec{\alpha}\cdot\vec{p}
+\beta\, m\, {\rm exp}
\left[i \vec{\hat{r}}\cdot\vec{\tau}\,
\gamma_5 \Theta(r)\right]\, .
\label{hedgehog}
\ee
The eigenstates $|\alpha\rangle$ of this Dirac Hamiltonian are in 
particular characterized by their grand--spin quantum number \cite{Ka84}. 
The grand--spin, $\vec{G}$ is the operator sum of total spin and 
isospin. Since $\vec{G}$ commutes with the Dirac Hamiltonian
(\ref{hedgehog}) the state ${\rm exp}(i\pi G_2)|\alpha\rangle$ is 
also an eigenstate of (\ref{hedgehog}) with energy $\epsilon_\alpha$ 
and grand--spin $G_\alpha$. This rotational symmetry, which actually
is a grand--spin reflection, will later be useful to simplify 
matrix elements of the quark wave--functions,~$\Psi_\alpha$.

For unit baryon number configurations it turns out that one distinct 
level, $\Psi_{\rm val}$, is strongly bound \cite{Al96}. This level
is referred to as the valence quark state. The total energy 
functional contains three pieces. The first one is due to the 
explicit occupation of the valence quark level to ensure unit 
baryon number. The second is the contribution of the polarized 
vacuum. It is extracted from the action (\ref{act1}) by considering 
an infinite time interval to discretize the eigenvalues of $\partial_t$. 
The sum over these eigenvalues then becomes a spectral integral 
\cite{Re89} which can be computed using Cauchy's theorem. Finally,  
there is the trivial part stemming from the local part of the 
action (\ref{act0}). Collecting these pieces we have \cite{Do92,Al96}
\be
E_{\rm tot}[\Theta]&=&
\frac{N_C}{2}\left(1-{\rm sign}(\epsilon_{\rm val})\right)
\epsilon_{\rm val}
-\frac{N_C}{2}\sum_{i=0}^2 c_i \sum_\alpha
\left\{\sqrt{\epsilon_\alpha^2+\Lambda_i^2}
-\sqrt{\epsilon_\alpha^{(0)2}+\Lambda_i^2}
\right\}
\nonumber \\ && \hspace{2cm}
+m_\pi^2f_\pi^2\int d^3r \, (1-{\rm cos}(\Theta))\, .
\label{etot}
\ee
Here we have also subtracted the vacuum energy associated
with the trivial meson field configuration and made use 
of the expressions obtained for $m_\pi$ and $f_\pi$ in the 
preceding subsection. The soliton is then obtained as the profile 
function $\Theta(r)$ which minimizes the total energy $E_{\rm tot}$
self--consistently.

At this point we have constructed a state which has unit baryon 
number but neither good quantum numbers for spin and flavor. Such 
states are generated by canonically quantizing the time--dependent
collective coordinates $A(t)$ which parameterize
the spin--flavor orientation of the soliton.
For a rigidly rotating soliton the Dirac operator 
becomes, after transforming to the flavor rotating
frame\footnote{A generalization to three flavors 
proceeds analogously but also requires to include
flavor symmetry breaking~\cite{We92}.} \cite{Re89},
\be
i\bDp=A\beta\left(i\partial_t - \tauom
-h\right)A^\dagger
\quad{\rm and}\quad
i\bDp_5=A\left(-i\partial_t + \tauom
-h\right)\beta A^\dagger\, .
\label{collq1}
\ee
Actual computations involve an expansion with respect to
the angular velocities 
\be
A^\dagger \frac{d}{dt} A = \frac{i}{2}\tauom\, .
\label{collq2}
\ee
According to the quantization rules, the angular velocities
are replaced by the spin operator 
\be
\vec{\Omega}\longrightarrow \frac{1}{\alpha^2}\, \vec{J}\, .
\label{collq3}
\ee
The constant of proportionality is the moment of inertia 
$\alpha^2$ which is calculated as a functional of the 
soliton \cite{Re89}. For the present purpose we remark that 
$\alpha^2$ is of the order $1/N_C$.\footnote{When formally considering 
nucleon structure functions it will turn out that it is mostly 
sufficient to refer to the Dirac operators as defined in 
eqs (\ref{collq1}) rather than to explicitly carry out the 
expansion in the angular velocities.} Hence an 
expansion in $\vec{\Omega}$ is equivalent to one 
in $1/N_C$. The nucleon wave--function becomes a (Wigner D) 
function of the collective coordinates. A useful relation in 
computing matrix elements of nucleon states is \cite{ANW}
\be
\langle N |\frac{1}{2}{\rm tr}
\left(A^\dagger\tau_i A\tau_j\right) |N\rangle =
-\frac{4}{3}\langle N | I_i J_j | N\rangle\, .
\label{collq4}
\ee

\bigskip
\stepcounter{chapter}
\leftline{\Large\it 3. A reminder on the pion structure function}
\medskip

Here we will reconsider the computation of the pion structure 
function \cite{Da95}. The reason is that in the present
formulation of the model this calculation goes beyond 
considering the standard `handbag' diagram because dimension 
five operators are involved. The matrix elements of
these operators will lead to isospin breaking contributions.
The particular way these contributions cancel among each other
will teach us how to organize the soliton calculation 
efficiently. 

\begin{figure}[tb]
\caption{\label{fig3_1}\sf Local diagrams which do not contribute to 
the absorptive part of the Compton amplitude in the Bjorken
limit. Dotted and curly lines refer to the pion and photon
fields, respectively.}
~
\centerline{
\epsfig{figure=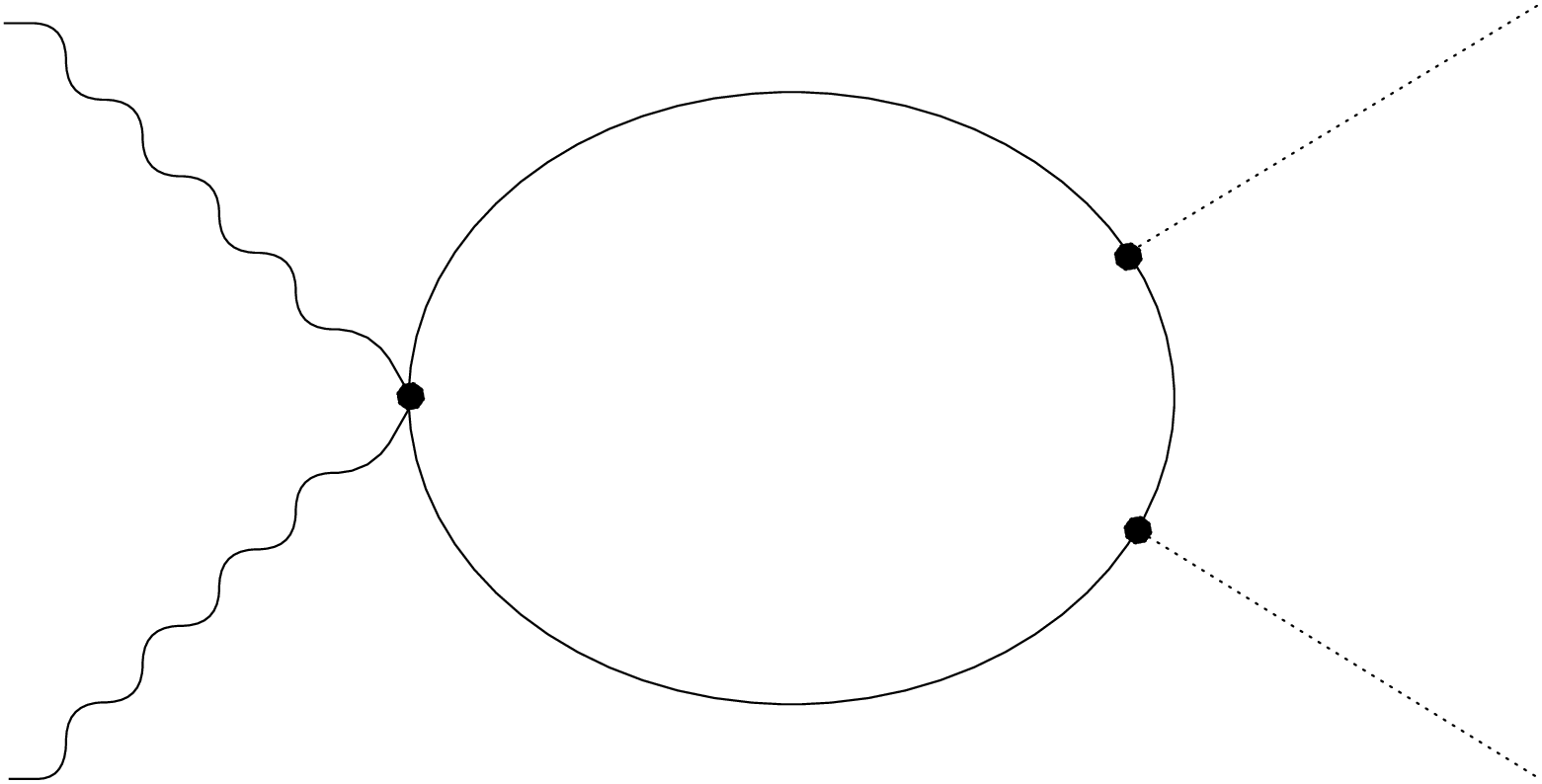,height=5cm,width=6.5cm}
\hspace{2cm}
\epsfig{figure=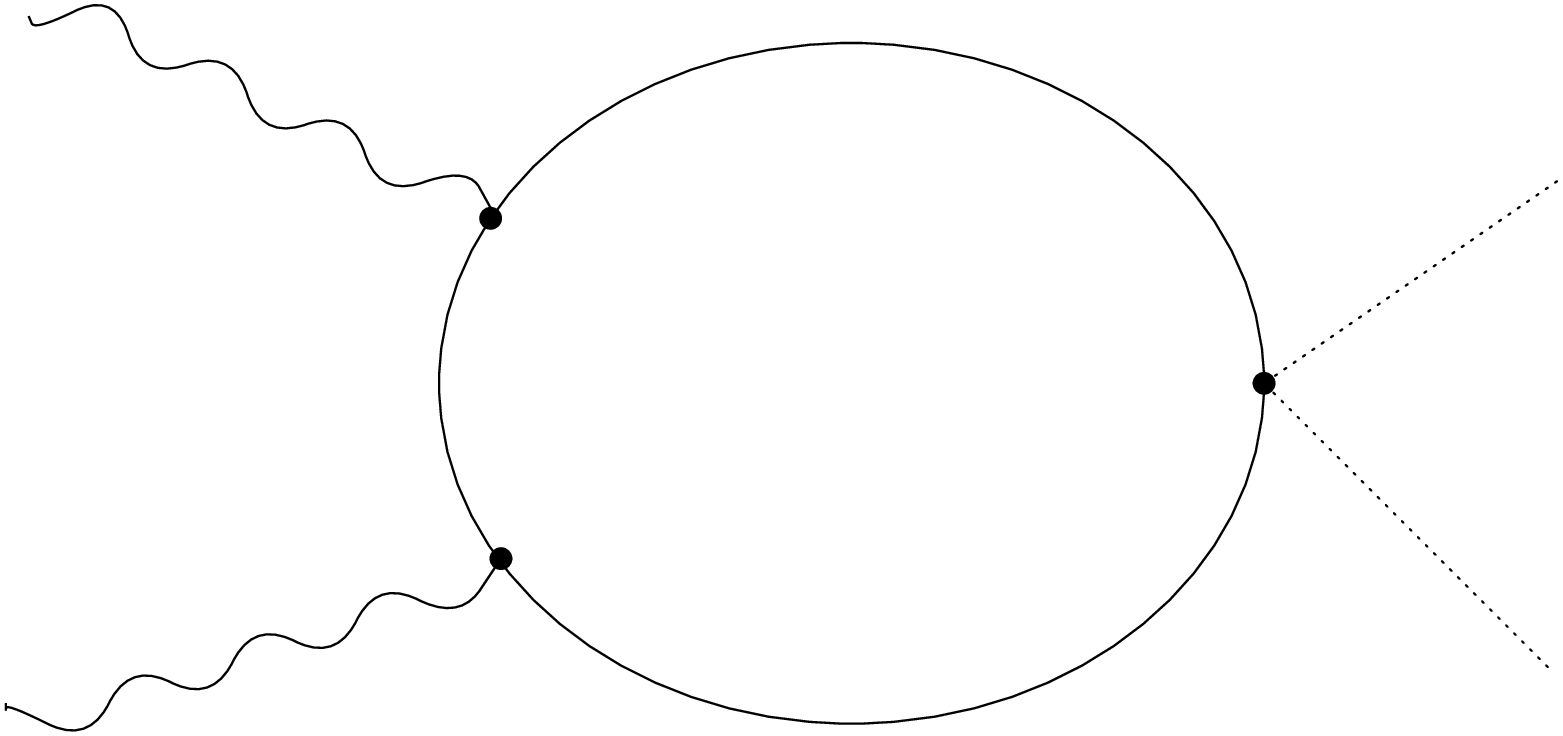,height=5cm,width=6.5cm}}
\end{figure}

\begin{figure}[htb]
\caption{\label{fig3_2}\sf Cutting the quark loop according
to Cutkosky's rules. External pion and photon lines are indicated
by dotted and curly lines. These diagrams have an isospin
violating ordering $(v_\mu\vec{\pi}v_\nu\vec{\pi})$.}
~
\centerline{
\epsfig{figure=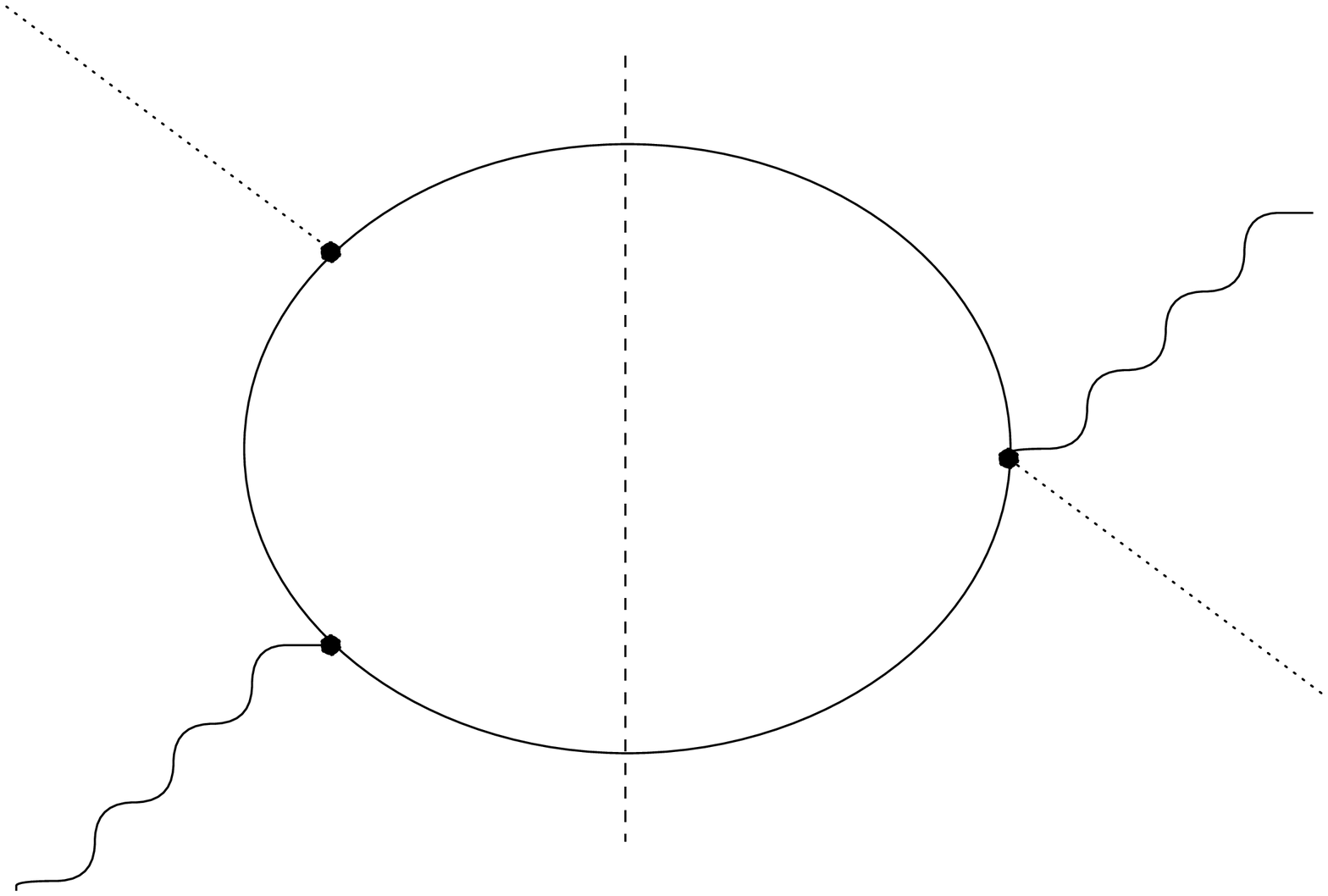,height=5cm,width=6.5cm}
\hspace{1cm}
\epsfig{figure=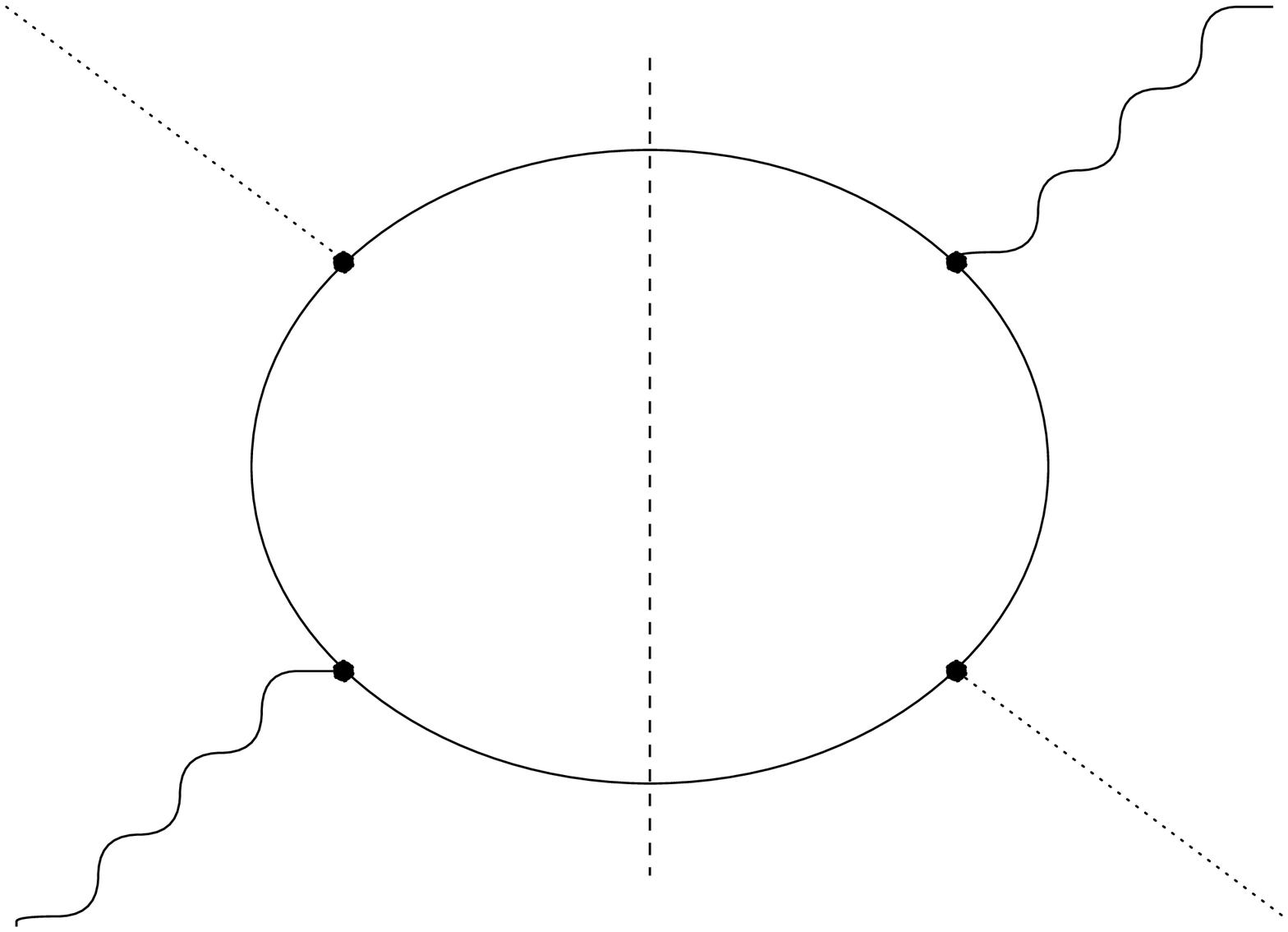,height=5cm,width=6.5cm}}
\end{figure}

\begin{figure}[htb]
\caption{\label{fig3_3}\sf Cutting the quark loop according
to Cutkosky's rules. External pion and photon lines are indicated
by dotted and curly lines, respectively. These diagrams have the 
proper isospin ordering $(v_\mu v_\nu\vec{\pi}\vec{\pi})$.}
~
\centerline{
\epsfig{figure=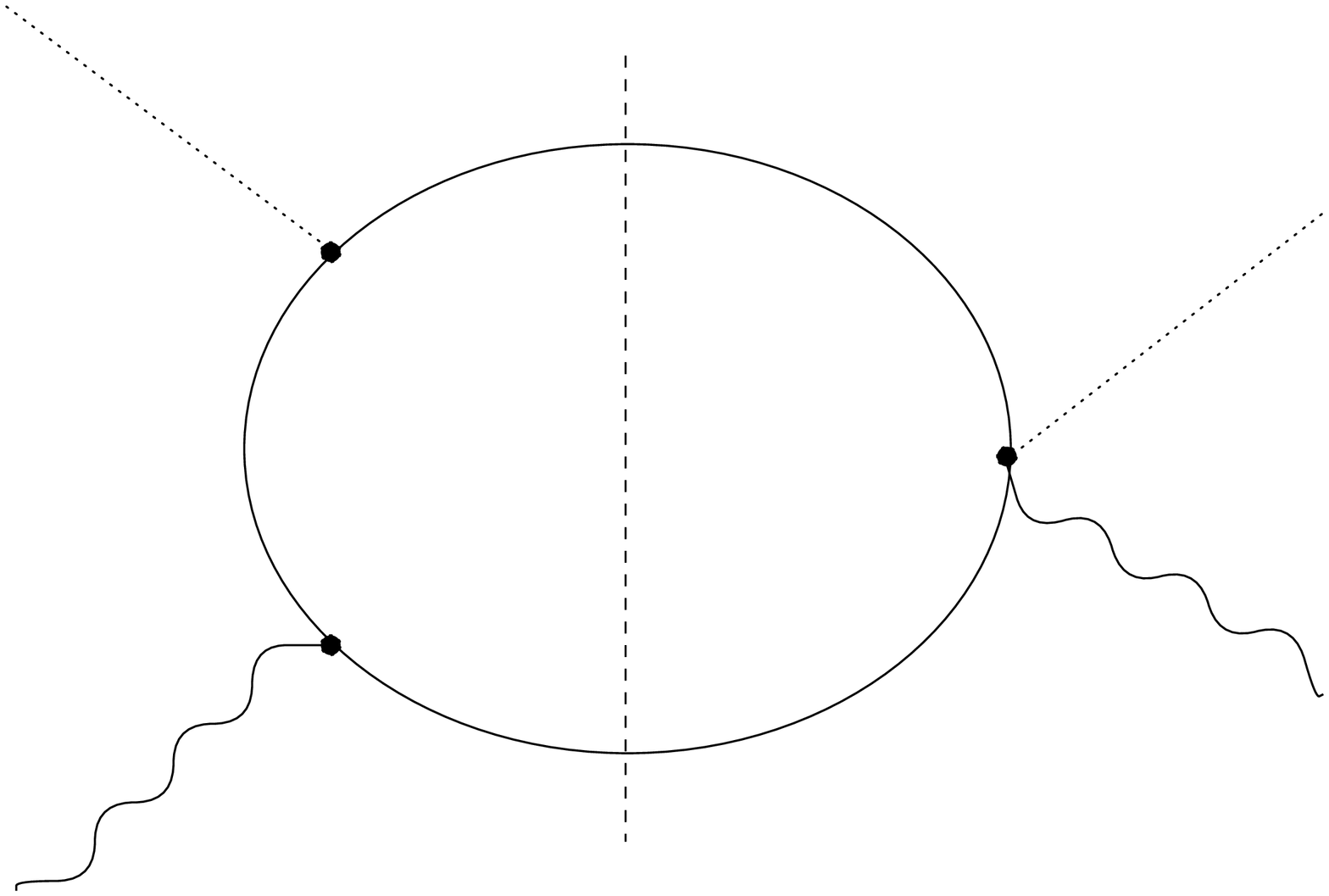,height=5cm,width=6.5cm}
\hspace{1cm}
\epsfig{figure=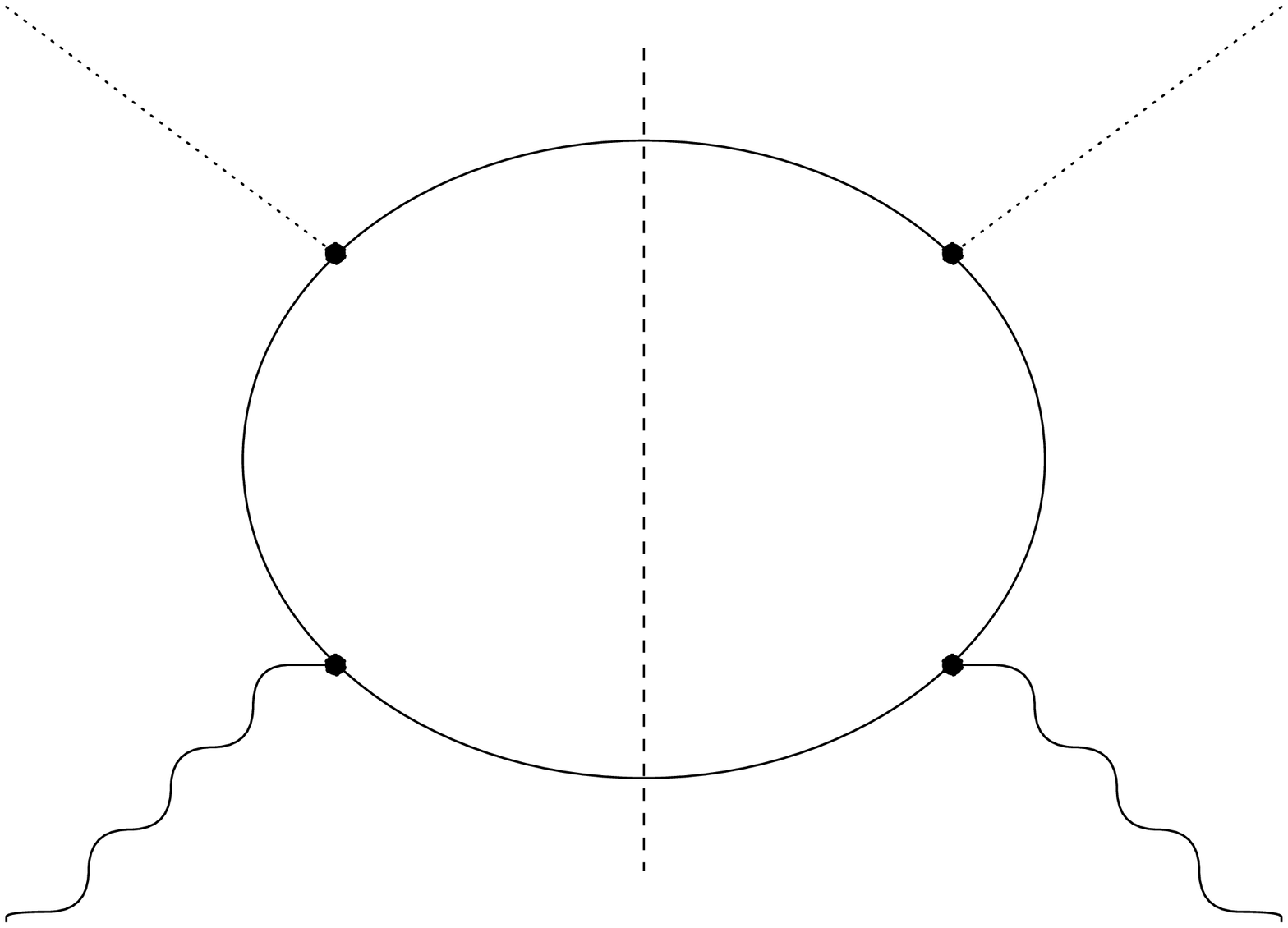,height=5cm,width=6.5cm}}
\end{figure}

DIS off pions is characterized by a single structure function,
$F(x)$,
\be
\frac{1}{2\pi} {\rm Im}\, T_{\mu\nu}(p,q) \quad \bjlim \quad
F(x)\left[-g_{\mu\nu}+\frac{q_\mu q_\nu}{q^2}
-\frac{1}{q^2}\left(p_\mu-\frac{q_\mu}{2x}\right)
\left(p_\nu-\frac{q_\nu}{2x}\right)\right]\, ,
\label{disp1}
\ee
where the Bjorken limit (\ref{bjl}) has been indicated.
In order to compute the Compton amplitude (\ref{disp1}) we 
calculate the time--ordered product
\be
T\left\{J_\mu(\xi) J_\nu(0)\right\}=
\frac{\delta^2}{\delta v^\mu(\xi)\delta v^\nu(0)} 
\A_{\rm NJL}[\vslash{\cal Q}]\Bigg|_{v_\mu=0}
\label{disp2}
\ee
from the action, $\A_{\rm NJL}$. Here ${\cal Q}$ represents the 
quark charge matrix. The vector (photon) field $v_\mu$ is 
added according to eqs (\ref{defd}) and (\ref{defd5}) while 
there are no axial sources. In order to compute the pion 
matrix element of (\ref{disp2}) we have to expand $\A_{\rm NJL}$ 
up to quadratic order in both the pion and photon fields. 
We note that diagrams like those in figure~\ref{fig3_1}, which
have identical particles at a single vertex, have no 
absorptive part in the Bjorken limit since these diagrams depend
on only one of the two external momenta. It is therefore
sufficient to consider
\be
-\bD\bD_5=\partial^2+m^2
+g\gamma_5\left[\dslash,\vec{\pi}\cdot{\tau}\right]
-i\left(\dslash\vslash{\cal Q}+\vslash{\cal Q}\dslash\right)
+ig\gamma_5\left[\vec{\pi}\cdot{\tau},\vslash{\cal Q}\right]+\ldots\, .
\label{disp3}
\ee
This expression shows that in the present formulation we are 
dealing with scalar quarks whose derivative interactions exhibit 
the spinorial nature of the quarks.
\noindent
Consequently, the substitution of (\ref{disp3}) into $\A_{\rm R}$ 
generates many diagrams. In particular we recognize from 
eq (\ref{disp3}) that the present formulation of the NJL model 
contains vertices which are not expected within a simple Dirac 
theory and we will have to go beyond computing standard 
`handbag' diagrams. First, there are pion--photon contact 
interactions contributing to diagrams like the one in the
left panel of figure~\ref{fig3_2}. In addition there 
are derivative couplings of the pions to the quarks. They
are dimension five operators which cause diagrams like 
that in the right panel of figure~\ref{fig3_2} not to vanish
in the Bjorken limit.
The diagrams in figure~\ref{fig3_2} have in common the 
$(\vec{\pi}v_\mu\vec{\pi}v_\nu)$ ordering of the pion and photon 
insertions.  The isospin part of the functional trace provides the 
factor ${\rm tr}\left(\tau_a{\cal Q}\tau_a{\cal Q}\right)$ where $a$ 
is the isospin quantum number of the pion. Apparently this leads to 
isospin violation as the contribution to the structure functions 
of charged and uncharged pions will be different. Fortunately
it turns out that the sum of all diagrams with this ordering
vanishes in the Bjorken limit. In addition there are diagrams with 
the vertex ordering $(v_\mu\vec{\pi}\vec{\pi}v_\nu)$. The respective 
isospin factors ${\rm tr}\left(\tau_a\tau_a{\cal Q}{\cal Q}\right)=
\frac{10}{9}$ do not depend on the isospin of the pion. 
Typical examples are depicted in figure~\ref{fig3_3}.
Collecting the contributions of all relevant graphs we find
\be
&&\langle \pi(p) |\frac{\delta^2}{\delta v^\mu(\xi)\delta v^\nu(0)}
\A_{\rm NJL}[\vslash{\cal Q}]\Bigg|_{v_\mu=0}|\pi(p) \rangle=
\nonumber \\* && \hspace{1cm}
\frac{5g^2N_C}{18}\sum_{i=0}^2c_i\int \frac{d\,^4k}{(2\pi)^4}
\frac{1}{-k^2+m^2+\Lambda_i^2-i\epsilon}\,
\frac{1}{\left[-(k-p)^2+m^2+\Lambda_i^2-i\epsilon\right]^2}
\nonumber \\ &&\hspace{2cm} \times
\Bigg\{\frac{-(k-p)^2+m^2+\Lambda_i^2}
{-(k-q-p)^2+m^2+\Lambda_i^2-i\epsilon}
{\rm tr}\left(\pslash\gamma^\mu\qslash\gamma^\nu
+\pslash\gamma^\nu\qslash\gamma^\mu\right)
\nonumber \\* && \hspace{2.5cm}
+\frac{-(k-p)^2+m^2+\Lambda_i^2}
{-(k+q-p)^2+m^2+\Lambda_i^2-i\epsilon}
{\rm tr}\left(\pslash\gamma^\mu\qslash\gamma^\nu
+\pslash\gamma^\nu\qslash\gamma^\mu\right)
\nonumber \\* && \hspace{2.5cm}
+2m^2\Bigg[\frac{{\rm tr}\left([\kslash-\pslash]
\gamma^\mu\qslash\gamma^\nu\right)}
{-(k+q-p)^2+m^2+\Lambda_i^2-i\epsilon}
\nonumber \\* && \hspace{5.0cm}
-\frac{{\rm tr}\left([\kslash-\pslash]\gamma^\nu
\qslash\gamma^\mu\right)}
{-(k-q-p)^2+m^2+\Lambda_i^2-i\epsilon}
\Bigg]\Bigg\}\, .
\label{disp4}
\ee
This expression simplifies considerably when applying
Cutkosky's rules in the Bjorken limit,
\be
\frac{1}{-k^2+m^2+\Lambda_i^2-i\epsilon}&&
\longrightarrow \quad -2i \pi \delta\left(k^2-m^2-\Lambda_i^2\right)\, ,
\label{disp5a}\\
\frac{1}{-(k\pm q-p)^2+m^2+\Lambda_i^2-i\epsilon}&&
\longrightarrow \quad -\frac{i \pi}{q^-} 
\delta\left(q^+\pm(k-p)^+\right)\, .
\label{disp5b}
\ee
Here we have already employed light--cone coordinates because they 
render the implementation of the Bjorken limit quite transparent; 
$q^-\to\infty$ and $q^+\to x p^+$. At this point we
pause to have a second look at the replacement (\ref{disp5b}).
In the $\delta$--function all terms which did not 
multiply $q^-$ have dropped out in the Bjorken limit.
In particular, all the dependence on the cut--offs, 
$\Lambda_i$, has disappeared. This yields the 
desired scaling behavior which would not have been recovered
in the frequently adopted proper--time regularization \cite{Da95}. 
For scaling to occur, a cut--off which is additive to the
quark mass appears mandatory.

The pion structure function is now straightforwardly obtained from 
$T_{11}+T_{22}$,
\be
F(x)&=&\frac{5}{9}(4 N_C g^2)\sum_{i=0}^2c_i
\int \frac{d\,^4k}{(2\pi)^4}
\frac{2\pi\delta\left(k^2-m^2-\Lambda_i^2\right)}
{\left[-(k-p)^2+m^2+\Lambda_i^2-i\epsilon\right]^2}
\nonumber \\* &&\hspace{1cm} \times
\Bigg\{\left[-(k-p)^2+m^2+\Lambda_i^2\right]p^+
\left[\delta\left(k^+-p^+-q^+\right)
-\delta\left(k^+-p^++q^+\right)\right]
\nonumber \\* &&\hspace{2cm}
+m^2\left(k^+-p^+\right)
\left[\delta\left(k^+-p^+-q^+\right)
-\delta\left(k^+-p^++q^+\right)\right]\Bigg\}\, .
\label{disp6}
\ee
The remaining integrals can be performed yielding
\be
F(x)&=&\frac{10}{9}N_Cg^2\sum_{i=0}^2c_i
\int \frac{d^2k_\bot}{(2\pi)^3}\frac{1}{x(1-x)}\,
\frac{M^2_i}{\left[M^2_i-m^2\right]^2}\, ,
\label{disp7}\\*
M^2_i&=&\frac{1}{x(1-x)}
\left[k_\bot^2+m^2+\Lambda_i^2\right]\, .
\nonumber 
\ee
Comparing this expression with the definition of the pion 
polarization function (\ref{specfct}) it can easily be verified 
that
\be
F(x)=\frac{5}{9} (4N_C g^2)
\frac{d}{dp^2}\left[p^2\Pi(p^2,x)\right]\Bigg|_{p^2=m_\pi^2}\, .
\label{disp8}
\ee
This well--known result \cite{Da95,Fr94} particularly implies 
that in the chiral limit ($m_\pi=0$) the pion structure function
equals unity, up to charge normalization. In appendix A we show 
how this result can also be obtained from the Bethe--Salpeter
amplitude (\ref{specfct}).

\bigskip
\stepcounter{chapter}
\leftline{\Large\it 4. Bjorken limit and scaling}
\medskip

\begin{figure}[htb]
\caption{\label{fig4_1}\sf Cutting the quark loop. The
thick lines denote quark propagators which contain the
full dependence on the pion fields. As in
figure \protect\ref{fig3_2} thin lines refer to free quark
propagators carrying the infinite photon momentum $q$.
The arrow indicates the Bjorken limit.}
~
\vskip0.5cm
\centerline{
\epsfig{figure=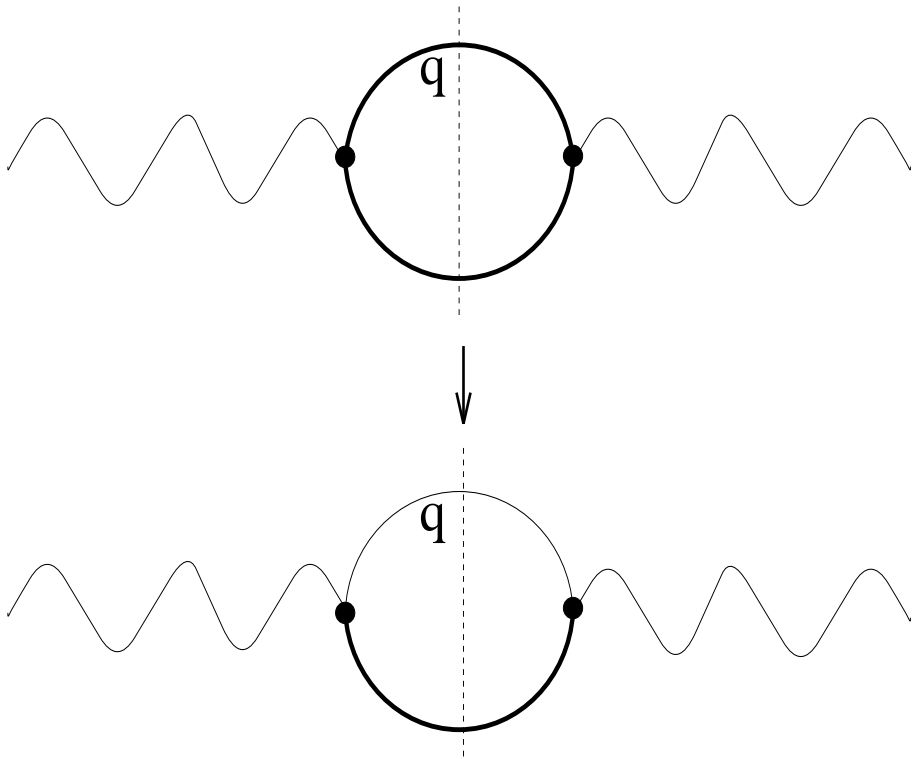,height=10cm,width=12cm}}
\end{figure}

\begin{figure}[hbt]
\caption{\label{fig4_2}\sf The first order perturbation to the lower
part of figure~\protect\ref{fig4_1} in expanding with respect to
the pion field which is denoted by the dotted line.}
~
\centerline{
\epsfig{figure=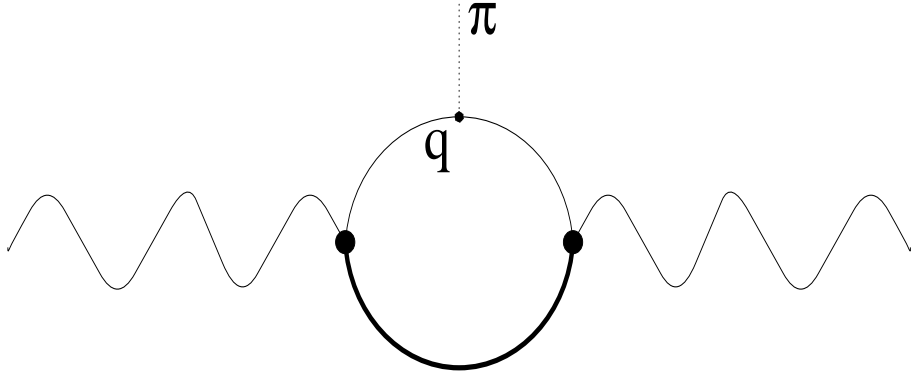,height=5cm,width=12cm}}
\end{figure}

At this point we recognize that the same result could have
been obtained by ignoring the issue of regularization and
simply computing the `handbag' diagram (right panel of 
figure~\ref{fig3_3}) using the vertices contained in
the simple Dirac operator,
\be
i\dslash + m +i g \gamma_5 \vec{\pi}\cdot\vec{\tau}
+\vslash{\cal Q} 
\nonumber 
\ee
and subsequently enforcing the Pauli--Villars regularization 
on the polarization function $\Pi(q^2)$. 
Hence, the question arises whether a simple argument exists
for the cancellation of all isospin violating diagrams, which
may involve dimension five operators, in the current approach.
The appearance of these terms actually is an artifact
of the simultaneous expansion in the pion and photon fields
(\ref{disp3}). According to eqs (\ref{defd}) and (\ref{defd5}) we
might equally well have expanded only in the photon field first
(omitting for the time being the trivial factors of the charge
matrix)
\be
-{\rm Tr}\,\left\{\left(-\bDp\bDp_5+\Lambda_i^2\right)^{-1}
\left[\left(\bDp\vslash+\vslash\bDp_5\right)
\left(-\bDp\bDp_5+\Lambda_i^2\right)^{-1}
\left(\bDp\vslash+\vslash\bDp_5\right)\right]\right\}\, .
\label{simple1}
\ee
Here square brackets have been introduced to mark those
parts through which the large photon momentum runs. Due to 
the cyclic properties of the trace this is merely
a choice. In momentum space the propagator inside the 
square brackets behaves like $1/Q^2$ in the Bjorken limit. In 
particular this implies that 
\be
\Big[\ldots\Big] &\longrightarrow&
\left(\bDp\vslash+\vslash\bDp_5\right)
\left(-\bDp\bDp_5\right)^{-1}
\left(\bDp\vslash+\vslash\bDp_5\right)
\nonumber \\ &\longrightarrow&
-\bDp\vslash\left(\bDp_5\right)^{-1}\vslash
-\vslash\left(\bDp\right)^{-1}\vslash\bDp_5\, .
\label{simple2}
\ee
This replacement tells us that the propagator through which 
the large photon momentum runs will not be effected by the 
regularization. Hence in the Bjorken limit there will be no 
contributions which behave like $Q^2/\Lambda_i^2$; thereby the
proper scaling behavior is manifest. In other regularization
schemes, like {\it e.g.} proper--time, wherein the cut--off is
not additive to the loop momenta, the absence of such scaling
violating contributions is not apparent. In the above expression
we have also omitted terms which contained a two photon vertex or 
were suppressed by additional factors of $1/Q^2$. Previously we 
expanded the operator (\ref{simple1}) in powers of the pion field 
leading to complicated three and four vertex quark loops as in 
figure~\ref{fig3_2}. Now we see that the Bjorken limit 
enforces the above observed cancellations among those 
diagrams. The expression (\ref{simple2}) simplifies even 
further by noting that the quark propagator inside the photon 
insertions carries the large photon momentum and should hence be 
approximated by the free massless propagator, 
\be
\Big[\ldots\Big] &\longrightarrow&
\bDp\vslash\left(\dslash\right)^{-1}\vslash
-\vslash\left(\dslash\right)^{-1}\vslash\bDp_5\, .
\label{simple3}
\ee
Together with the propagator in (\ref{simple1}) this will yield
the standard `handbag' diagram with the propagators connecting
the quark--pion and quark--photon vertices regularized according 
to the Pauli--Villars scheme. The transition from the expression 
(\ref{simple1}) to (\ref{simple3}) is illustrated in 
figure~\ref{fig4_1}.
One may still wonder whether the separation into (infinitely) large 
and small momentum components of the expression (\ref{simple1}) 
would be arbitrary, after all the functional trace implies the 
integration over all momenta. For this purpose assume that
the momenta flowing through the two propagators in the
functional trace (\ref{simple1}) would both be comparable to
the photon momentum $Q^2$. Then also the propagator 
outside the square brackets can be approximated by
the free massless propagator and the object (\ref{simple1})
simplifies even further,
\be
2{\rm Tr}\,\left\{\left(\dslash\right)^{-1}\vslash
\left(\dslash\right)^{-1}\vslash\right\}\, .
\label{simple4}
\ee
This will contribute a disconnected graph whose $q$--dependence 
is completely given by
\be
\left(2q^\mu q^\nu-q^2g^{\mu\nu}\right)\, 
\frac{1}{q^2+i\epsilon}\, \frac{1}{q^2+i\epsilon}\, .
\label{simple5}
\ee
This contribution vanishes like $1/q^2$ in the Bjorken 
limit. The fact that the external momenta (pion or nucleon 
versus photon) are on completely different scales allows us to 
distinguish the quark propagators accordingly. We could furthermore 
ask for possible corrections to the expression (\ref{simple3}) 
when taking the quark propagator carrying $q$ not 
to be the free massless one. These corrections would come about as
an expansion in the pion field contained in the Dirac
operators $\bDp$ and $\bDp_5$. The first order correction
is shown in figure~\ref{fig4_2}.
As a matter of fact this graph is similar to the one in
the right panel of figure~\ref{fig3_2} with non--trivial
quark propagators on either side of the photon vertices.
Once we have organized the expansion such that the dimension--five
operators no longer contribute (at the end they
should cancel anyhow), such diagrams will be suppressed
in the Bjorken limit. Thus the replacement in (\ref{simple3}) 
is indeed exact in the Bjorken limit. In appendix B we discuss
this replacement in the context of the Wigner transformation.

Having clarified these issues for the pion structure functions we 
now turn to the case of the nucleon. After re--introducing the charge 
matrix factors it will be sufficient to differentiate
\be
\A_{\Lambda,{\rm R}}^{(2,v)}=
-i\frac{N_C}{4}\sum_{i=0}^2c_i
{\rm Tr}\,\left\{\left(-\bDp\bDp_5+\Lambda_i^2\right)^{-1}
\left[{\cal Q}^2\vslash\left(\dslash\right)^{-1}\vslash\bDp_5
-\bDp(\vslash\left(\dslash\right)^{-1}\vslash)_5
{\cal Q}^2\right]\right\}
\label{simple6}
\ee
with respect to the vector sources in order to obtain the Compton 
amplitude in the Bjorken limit. 

At this point it is important to explain the crucial role of the 
subscript `5' attached to the second term in square brackets 
of eq (\ref{simple6}). For this second term we have to recall 
that the (inverse) derivative operator in 
$\vslash\left(\dslash\right)^{-1}\vslash$ is actually 
associated with the expansion of $\bD_5$. When comparing this 
$\gamma_5$--odd operator to the ordinary Dirac operator in eqs 
(\ref{defd}) and (\ref{defd5}) one observes immediately that $\bD_5$
has a relative sign between the derivative operator $i\dslash$ and 
the axial source $\aslash\gamma_5$. In section 6 we will show that
for the regularization of the
structure functions to be consistent with the sum rules this relative 
sign must also be reflected in the Dirac decomposition of 
$(\vslash\left(\dslash\right)^{-1}\vslash)_5$. Therefore the 
axial--vector component of 
$(\vslash\left(\dslash\right)^{-1}\vslash)_5$ 
requires a relative sign.
With $S_{\mu\rho\nu\sigma}=g_{\mu\rho}g_{\nu\sigma}
+g_{\rho\nu}g_{\mu\sigma}-g_{\mu\nu}g_{\rho\sigma}$, that is
\be
\gamma_\mu\gamma_\rho\gamma_\nu
=S_{\mu\rho\nu\sigma}\gamma^\sigma
-i\epsilon_{\mu\rho\nu\sigma}\gamma^\sigma\gamma^5
\quad {\rm while} \quad
(\gamma_\mu\gamma_\rho\gamma_\nu)_5
=S_{\mu\rho\nu\sigma}\gamma^\sigma+
i\epsilon_{\mu\rho\nu\sigma}\gamma^\sigma\gamma^5\, .
\label{defsign}
\ee
The fact that the sum rules enforce this extension of the 
regularization scheme is not all surprising when noting
that the derivative operator $i\dslash$ fixes the Noether 
currents. Rather it is imposed and a consequence of the `sum rules' 
of the model defined by $\bD_5$, {\it cf.} appendix C. We recall 
that the $\bD_5$ model, which is not physical, has been 
introduced as a device to allow for a regularization which 
maintains the anomaly structure of the underlying theory. Hence it 
is not at all surprising that further specification of this 
regularization prescription is demanded in order to formulate a fully
consistent model. It should be stressed that this issue is
not specific to the Pauli--Villars scheme but rather all schemes 
which do regularize the sum, ${\rm log}\,(\bD)+{\rm log}\,(\bD_5)$
but not the difference, ${\rm log}\,(\bD)-{\rm log}\,(\bD_5)$
will require the specification (\ref{defsign}). Since only the 
polarized, {\it i.e.} spin dependent, structure functions are effected, 
this issue has not shown up when computing the pion structure functions 
in the Pauli--Villars scheme.

For the imaginary part of the action the expression analogous to
(\ref{simple6}) reads
\be
\A_{\Lambda,{\rm I}}^{(2,v)}=
-i\frac{N_C}{4}
{\rm Tr}\,\left\{\left(-\bDp\bDp_5\right)^{-1}
\left[{\cal Q}^2\vslash\left(\dslash\right)^{-1}\vslash\bDp_5
+\bDp(\vslash\left(\dslash\right)^{-1}\vslash)_5
{\cal Q}^2\right]\right\}\, .
\label{simple7}
\ee
Again, it is understood that the large photon momentum runs only 
through the operators in square brackets. 

It is also noticed that in the unregularized case ($\Lambda_i\equiv0$)
the contributions associated to the expansion of $\bD_5$ cancel in the sum 
(\ref{simple6}) and (\ref{simple7}) leaving
\be
\A^{(2,v)}=
i\frac{N_C}{2}
{\rm Tr}\,\left\{\left(\bDp\right)^{-1}
\left[{\cal Q}^2\vslash\left(\dslash\right)^{-1}
\vslash\right]\right\}\, .
\label{simple8}
\ee
This expression generates the `handbag' diagram upon expansion in 
the pion field. A similar cancellation would occur if the 
imaginary part were also regularized. In particular, the 
$\gamma_5$--odd pieces stemming from expanding $\bD_5$ would
cancel. With these remarks on the regularization of the Compton
tensor in the Bjorken limit we are prepared to proceed with
the full soliton calculation.

\bigskip
\stepcounter{chapter}
\leftline{\Large\it 5. The nucleon structure functions}
\medskip

Here we will only discuss the contribution of the 
polarized vacuum to the nucleon structure functions.
The contribution of the distinct valence level, which
is not effected by the regularization, {\it cf.}
eq (\ref{etot}), has previously been detailed \cite{Ja75,We96}.

Before carrying out a calculation like that in section 3 we 
must define the hadronic tensor for localized field configurations. 
Among other things this demands the restoration of translational 
invariance. This is accomplished by introducing a collective 
coordinate, $\vec{R}$, which describes the position of a 
soliton (nucleon) \cite{Ge76} with its momentum, $\vec{p}$ 
being conjugate to this collective coordinate, {\it i.e.} 
$\langle \vec{R}|\vec{p}\,\rangle=
\sqrt{2E}\,{\rm exp}\left(i\vec{R}\cdot\vec{p}\right)$.
Here $E=\sqrt{\vec{p}\,^2+M_N^2}$ denotes the nucleon 
energy. The Compton amplitude (\ref{Comp2}) is then obtained
by taking the pertinent matrix element and averaging over the 
position of the soliton,
\be
T_{\mu\nu}^{ab} &=& 2i M_N
\int d^4\xi\, \int d^3R\, e^{iq\cdot \xi}\,
\Big\langle p,s\Big| T\left\{J_{\mu}^{a}(\xi-R)
J_{\nu}^{b \dagger}(-R)\right\}\Big| p,s \Big\rangle
\nonumber\\
&=& 2i M_N
\int d^4\xi_1\, \int d^3\xi_2\, e^{iq\cdot(\xi_1-\xi_2)}
\Big\langle s\Big| T\left\{J_{\mu}^{a}(\xi_1)
J_{\nu}^{b \dagger}(\xi_2)\right\}\Big| s \Big\rangle\, .
\label{nuc1}
\ee
Here we have made use of the fact that the 
initial and final nucleon states have identical momenta. 
Furthermore we will treat $\xi_2$ as a four--vector 
keeping in mind that its temporal component vanishes,
$\xi_2^0=0$. Finally the spin--flavor matrix elements will 
be evaluated in the space of the collective coordinates $A$, 
which have been introduced in eq (\ref{collq1}).

\bigskip
\leftline{\large\it 5a. The unregularized case}
\medskip
Before going into details we wish to briefly discuss the 
formal result associated with the unregularized case (\ref{simple8})
as a warm--up. According to eq (\ref{nuc1}) we evaluate 
\be
T\left(J_\mu(\xi_1),J_\nu(\xi_2)\right)
&=&iN_C \frac{\delta^2}
{\delta v^\mu(\xi_1)\delta v^\nu(\xi_2)}\,
{\rm Tr}\,\left\{\left(-\bDp\right)^{-1}
\left[{\cal Q}^2\vslash\left(\dslash\right)^{-1}
\vslash\right]\right\}\Bigg|_{v_\mu=0}
\hspace{2.5cm}~
\nonumber \\ 
&=&i\frac{N_C}{2}\, {\rm Tr}\,
\Bigg\{\left(-\bDp\right)^{-1}{\cal Q}^2
\Big[\gamma_\mu\delta^4\left(\hat{x}-\xi_1\right)
\left(\dslash\right)^{-1}
\gamma_\nu\delta^4\left(\hat{x}-\xi_2\right)
\nonumber \\ && \hspace{5cm}
\,+\gamma_\nu\delta^4\left(\hat{x}-\xi_2\right)
\left(\dslash\right)^{-1}
\gamma_\mu\delta^4\left(\hat{x}-\xi_1\right)\Big]\Bigg\}\, .
\label{nuc2}
\ee
Here $\hat{x}$ refers to the position operator. Its
introduction represents a suitable tool to evaluate the 
functional trace because $\hat{x}|x\rangle=x|x\rangle$.  This 
functional trace is computed by using a plane--wave
basis for the operator in the square brackets while 
the matrix elements of $\bDp$ are evaluated employing
the eigenfunctions $\Psi_\alpha$ of the Dirac Hamiltonian
(\ref{hedgehog}). Confining ourselves to the leading
order in $1/N_C$ piece, {\it i.e.} omitting, for the time
being, the corrections due to the collective rotations
yields
\be
T_{\mu\nu}(q)\hspace{-0.2cm}&=&\hspace{-0.2cm}
-M_NN_C\int\frac{d\omega}{2\pi} \sum_\alpha
\int d^4\xi_1\,\int d^3\xi_2\, \int \frac{d^4k}{(2\pi)^4}\,
{\rm e}^{i\xi_1^0(q^0+k^0)}\,
{\rm e}^{-i(\vec{\xi}_1-\vec{\xi}_2)\cdot(\vec{q}+\vec{k})}\,
\frac{1}{k^2+i\epsilon}\hspace{1cm}~
\nonumber \\ && \hspace{2.0cm}\times
\frac{\omega+\epsilon_\alpha}
{\omega^2-\epsilon^2_\alpha+i\epsilon}
\Big\langle N \Big|
\Bigg\{\bar{\Psi}_\alpha(\vec{\xi}_1){\cal Q}_A^2
\gamma_\mu \kslash\gamma_\nu\Psi_\alpha(\vec{\xi}_2)\,
{\rm e}^{i\xi_1^0\omega}
\nonumber \\ && \hspace{6.0cm}
-\,\bar{\Psi}_\alpha(\vec{\xi}_2){\cal Q}_A^2
\gamma_\nu \kslash\gamma_\nu\Psi_\alpha(\vec{\xi}_1)\,
{\rm e}^{-i\xi_1^0\omega}\Bigg\}\Big| N\Big\rangle\, .
\label{nuc3}
\ee
A few remarks are in order. First, the frequency integral 
stems from summing over the eigenvalues of $i\partial_t$ contained
in $\bDp$ in the limit of an infinitely large time interval
(in Euclidean space or with Feynman boundary conditions). This 
large time interval singles out the vacuum contribution to 
the functional trace \cite{Re89}. Second, the
momentum integral over $k$ essentially introduces the
Fourier transformation of the single quark wave--functions
$\Psi_\alpha$ which originated from the projection of these states
on the elements of the plane wave--basis. Note also that
the Dirac matrix $\beta$, contained in the definition 
(\ref{defh}) re--assembled to the adjoint spinor
$\bar{\Psi}_\alpha=\Psi^\dagger_\alpha\beta$. As we will see, 
this need not be the case in the fully regularized treatment. 
The collective coordinates, $A$ are only treated 
at the leading order in $1/N_C$ by rotating the charge
matrix ${\cal Q}^2 \to {\cal Q}^2_A=A^\dagger{\cal Q}^2A$ and 
taking the nucleon matrix elements in the space of the collective 
coordinates (\ref{collq4}). The next--to--leading order can 
straightforwardly be accounted for by expanding the matrix elements 
of the operator (\ref{collq2}),
\be
\langle\omega,\alpha|\left(\bDp\right)^{-1}|\omega,\beta\rangle
=\frac{\delta_{\alpha\beta}}{\omega-\epsilon_\alpha}
+\frac{1}{\omega-\epsilon_\alpha}\,
\langle \alpha|\tauom|\beta\rangle\,
\frac{1}{\omega-\epsilon_\beta} + 
{\cal O}\left(\vec{\Omega}\,^2\right)\, .
\label{nuc4}
\ee
In addition there will be $1/N_C$ corrections associated 
with the bi--locality in the time coordinates $\xi_1^0=t$
and $\xi_2^0=0$. This bi--locality not only effects the 
eigen--functions of $i\partial_t$ but also the collective 
coordinates. However, it can be rewritten as a local quantity 
by expanding \cite{Wa98,Po99,Wa98a}
\be
\langle t, \vec{\xi}| A(\hat{t}) | \omega,\alpha\rangle
&=& A(t)\, {\rm e}^{-i\omega t} \Psi_\alpha(\vec{\xi})
\nonumber \\ &=&
A(0)\left[1+\frac{it}{2}\,\tauom\right]
{\rm e}^{-i\omega t} \Psi_\alpha(\vec{\xi})+
{\cal O}\left(\vec{\Omega}\,^2\right)\, .
\label{nuc4a}
\ee
Apparently the non--linearity in the time--coordinate
becomes feasible in the large $N_C$ expansion since the 
angular velocity, $\vec{\Omega}$ is of order $1/N_C$, 
{\it cf.} eq (\ref{collq3}). We will see later that the 
contribution to the structure functions stemming from the 
$t$--dependent coefficient of the angular velocity in 
(\ref{nuc4a}) is most conveniently expressed as a derivative with 
respect to Bjorken $x$. As the induced terms (\ref{nuc4}) 
and (\ref{nuc4a})  do not depend on the photon momentum, $q$, 
their inclusion in the Compton amplitude is just a matter of 
book--keeping rather than a principle obstacle.

Putting the quark fields on--shell according to Cutkosky's rules 
identifies the imaginary part of the Compton amplitude and hence 
the hadronic tensor
\be
W_{\mu\nu}\hspace{-0.2cm}&=&\hspace{-0.2cm}
M_NN_C\sum_\alpha {\rm sign}(\epsilon_\alpha)
\int dt\, \int d^3\xi_1\,\int d^3\xi_2\, 
\int \frac{d^4k}{(2\pi)^4}\, {\rm e}^{it(q^0+k^0)}\,
{\rm e}^{-i(\vec{\xi}_1-\vec{\xi}_2)\cdot(\vec{q}+\vec{k})}\,
\delta\left(k^2\right)k^\rho \hspace{1cm}~
\nonumber \\* && \hspace{0.5cm}\times
\Big\langle N \Big|S_{\mu\rho\nu\sigma}
\Bigg\{\bar{\Psi}_\alpha(\vec{\xi}_1)
{\cal Q}_A^2\gamma^\sigma\Psi_\alpha(\vec{\xi}_2)\,
{\rm e}^{i\epsilon_\alpha t}
-\bar{\Psi}_\alpha(\vec{\xi}_2)
{\cal Q}_A^2\gamma^\sigma\Psi_\alpha(\vec{\xi}_1)\,
{\rm e}^{-i\epsilon_\alpha t}\Bigg\}
\nonumber \\* && \hspace{1.0cm}
-i\epsilon_{\mu\rho\nu\sigma}
\Bigg\{\bar{\Psi}_\alpha(\vec{\xi}_1)
{\cal Q}_A^2\gamma^\sigma\gamma_5\Psi_\alpha(\vec{\xi}_2)\,
{\rm e}^{i\epsilon_\alpha t}
+\bar{\Psi}_\alpha(\vec{\xi}_2)
{\cal Q}_A^2\gamma^\sigma\gamma_5\Psi_\alpha(\vec{\xi}_1)\,
{\rm e}^{-i\epsilon_\alpha t}\Bigg\}
\Big| N\Big\rangle\, .
\label{nuc5}
\ee
It is recognized that this expression for the hadronic tensor 
is analogous to the one obtained in the bag model some time 
ago \cite{Ja75}. For the purpose of comparison we have 
introduced the time coordinate $t=\xi_1^0$.
Although the nature of the quark spinors in 
the NJL and bag models is quite different, this formal
equality is not surprising because all that has 
entered so far, have been the properties of the symmetry
currents in the model. Moreover, in both models these
currents are formally identical to currents of a 
non--interacting Dirac theory. At this order in $1/N_C$ the 
only difference is contained in the labels ``$\alpha$''
which characterize the single quark levels.

We also recognize two different contributions to the hadronic 
tensor: one associated with forward moving intermediate quarks 
($\xi_1,\mu\to\xi_2,\nu$) and one wherein those quarks propagate 
backward ($\xi_2,\nu\to\xi_1,\mu$). In the parton model language 
these two pieces correspond to quark and anti--quark 
distributions\footnote{It should be recalled that presently the 
quark field operators refer to the degrees of freedom in
the NJL model. Hence such distributions may not necessarily
be identical to those of the QCD current quarks.}, respectively.

\bigskip
\leftline{\large\it 5b. The fully regularized case}
\medskip
Here we compute the contribution to the hadronic tensor stemming
from the time--ordered product in (\ref{simple6}). It is just
a matter of generalizing the previous calculation to obtain the 
contribution of the regularized real part, $\A_{\rm R}$ to the 
hadronic tensor. For the time being we will confine ourselves to the 
leading order contribution in $1/N_C$ while the resulting formulas 
including the cranking corrections are presented in appendix D. 
We find
\be
T^{({\rm R})}_{\mu\nu}&=&-M_N\frac{N_C}{2}
\int\frac{d\omega}{2\pi} 
\sum_\alpha \int dt\, \int d^3\xi_1\,\int d^3\xi_2\, 
\int \frac{d^4k}{(2\pi)^4}\, {\rm e}^{i(q^0+k^0)t}\,
{\rm e}^{-i(\vec{\xi}_1-\vec{\xi}_2)\cdot(\vec{q}+\vec{k})}\,
\frac{1}{k^2+i\epsilon}\hspace{1.2cm}~
\label{areg1} \\ && \hspace{-1.8cm}\times
\Big\langle N \Big|\sum_{i=0}^2c_i\Bigg\{
\frac{\omega+\epsilon_\alpha}
{\omega^2-\epsilon^2_\alpha-\Lambda^2_i+i\epsilon}
\left[\Psi^\dagger_\alpha(\vec{\xi}_1)\beta{\cal Q}_A^2\gamma_\mu
\kslash\gamma_\nu\Psi_\alpha(\vec{\xi}_2)\,
{\rm e}^{i\omega t}
-\,\Psi^\dagger_\alpha(\vec{\xi}_2)\beta{\cal Q}_A^2\gamma_\nu
\kslash\gamma_\mu\Psi_\alpha(\vec{\xi}_1)\,
{\rm e}^{-i\omega t}\right]
\nonumber \\ &&\hspace{-1.8cm}
+\, \frac{\omega-\epsilon_\alpha}
{\omega^2-\epsilon^2_\alpha-\Lambda^2_i+i\epsilon}
\left[\Psi^\dagger_\alpha(\vec{\xi}_1){\cal Q}_A^2
(\gamma_\mu\kslash\gamma_\nu)_5\beta\Psi_\alpha(\vec{\xi}_2)\,
{\rm e}^{i\omega t}
-\,\Psi^\dagger_\alpha(\vec{\xi}_2){\cal Q}_A^2
(\gamma_\nu\kslash\gamma_\mu)_5\beta\Psi_\alpha(\vec{\xi}_1)\,
{\rm e}^{-i\omega t}\right]\Bigg\}
\Big| N\rangle\, .
\nonumber
\ee
The first term in curly brackets corresponds the previous 
result (\ref{nuc3}) in the limit of vanishing regularization.
The second term, however, is new and originates from constructing
the real part by adding the $\bD_5$ model. It is also noticed 
that in this term the Dirac matrix $\beta$ does not combine to 
the adjoint spinor $\bar{\Psi}_\alpha$. Before applying Cutkosky's 
rules to (\ref{areg1}) in order to extract the hadronic tensor, we 
will compute the piece stemming from the time--ordered product in 
(\ref{simple7}) which is associated to the imaginary part of the 
bosonized NJL model action. Also in this case, the cranking corrections 
are relegated to appendix D. In leading order $1/N_C$ the imaginary part 
of the action contributes
\be
T^{({\rm I})}_{\mu\nu}&=&-M_N\frac{N_C}{2}
\int\frac{d\omega}{2\pi}
\sum_\alpha \int dt\, \int d^3\xi_1\,\int d^3\xi_2\,
\int \frac{d^4k}{(2\pi)^4}\, {\rm e}^{i(q^0+k^0)t}\,
{\rm e}^{-i(\vec{\xi}_1-\vec{\xi}_2)\cdot(\vec{q}+\vec{k})}\,
\frac{1}{k^2+i\epsilon}\hspace{1.2cm}~
\label{ireg1} \\ && \hspace{-1.0cm}\times
\Big\langle N\Big|\Bigg\{
\frac{\omega+\epsilon_\alpha}
{\omega^2-\epsilon^2_\alpha+i\epsilon}
\left[\Psi^\dagger_\alpha(\vec{\xi}_1)\beta{\cal Q}_A^2\gamma_\mu
\kslash\gamma_\nu\Psi_\alpha(\vec{\xi}_2)\,
{\rm e}^{i\omega t}
-\,\Psi^\dagger_\alpha(\vec{\xi}_2)\beta{\cal Q}_A^2\gamma_\nu
\kslash\gamma_\mu\Psi_\alpha(\vec{\xi}_1)\,
{\rm e}^{-i\omega t}\right]
\nonumber \\ &&\hspace{-1.0cm}
-\, \frac{\omega-\epsilon_\alpha}
{\omega^2-\epsilon^2_\alpha+i\epsilon}
\left[\Psi^\dagger_\alpha(\vec{\xi}_1){\cal Q}_A^2
(\gamma_\mu\kslash\gamma_\nu)_5\beta\Psi_\alpha(\vec{\xi}_2)\,
{\rm e}^{i\omega t}
-\,\Psi^\dagger_\alpha(\vec{\xi}_2){\cal Q}_A^2
(\gamma_\nu\kslash\gamma_\mu)_5\beta\Psi_\alpha(\vec{\xi}_1)\,
{\rm e}^{-i\omega t}\right]\Bigg\}
\Big| N\Big\rangle
\nonumber
\ee
to the Compton tensor. Except for the regularization this 
term differs from (\ref{areg1}) in the sign of the second term. 
Of course, if the regularization of 
$T_{\mu\nu}^{({\rm R})}$ were omitted, the sum of (\ref{areg1})
and (\ref{ireg1}) would combine to (\ref{nuc3}).

Now we can put pieces together providing the full Compton
amplitude in the Bjorken limit
\be
T_{\mu\nu}(q)&=&-M_N\frac{N_C}{2}\int \frac{d\omega}{2\pi}
\sum_\alpha \int dt\, \int d^3\xi_1\,\int d^3\xi_2\,
\int \frac{d^4k}{(2\pi)^4}\,
{\rm e}^{i(q_0+k_0)t}\,
{\rm e}^{-i({\vec q}+{\vec k})\cdot({\vec \xi}_1-{\vec \xi}_2)}\,
\frac{1}{k^2+i\epsilon}
\nonumber \\ && \hspace{-1cm}\times \Big\langle N\Big|
\Bigg\{\left[{\rm e}^{i\omega t}
\Psi_\alpha^\dagger({\vec \xi}_1)\beta {\cal Q}_A^2\gamma_\mu\kslash
\gamma_\nu \Psi_\alpha({\vec \xi}_2)
-{\rm e}^{-i\omega t}
\Psi_\alpha^\dagger({\vec \xi}_2)\beta {\cal Q}_A^2\gamma_\nu\kslash
\gamma_\mu \Psi_\alpha({\vec \xi}_1)\right]f_\alpha^+(\omega)
\label{treg} \\ &&\hspace{-0.5cm}
+\left[{\rm e}^{i\omega t}
\Psi_\alpha^\dagger({\vec \xi}_1){\cal Q}_A^2
(\gamma_\mu\kslash\gamma_\nu)_5\beta 
\Psi_\alpha({\vec \xi}_2)-{\rm e}^{-i\omega t}
\Psi_\alpha^\dagger({\vec \xi}_2){\cal Q}_A^2
(\gamma_\nu\kslash\gamma_\mu)_5\beta 
\Psi_\alpha({\vec \xi}_1)\right]f_\alpha^-(\omega)
\Bigg\}\Big| N\Big\rangle\, ,
\nonumber
\ee
with the spectral functions
\be
f_\alpha^\pm(\omega)=\sum_{i=0}^2 c_i \frac{\omega\pm\epsilon_\alpha}
{\omega^2-\epsilon_\alpha^2-\Lambda_i^2+i\epsilon}
\pm\frac{\omega\pm\epsilon_\alpha}
{\omega^2-\epsilon_\alpha^2+i\epsilon}\, .
\label{sfct}
\ee
It is clear that in case we had chosen to regularize $\A_{\rm I}$ 
similarly to $\A_{\rm R}$ the terms involving $f_\alpha^-(\omega)$ 
would disappear and consequently the prescription (\ref{defsign}),
associated with the spin dependent part of $\bD_5$, would not
have been required.

As the quark wave--functions are localized, the coordinate space 
integrals gather most of their support when 
${\xi}_\sigma^1\sim\xi_\sigma^2$. Hence only that region of the 
$k$--integral will be relevant which has $k_\sigma\sim -q_\sigma$, 
{\it i.e.} the loop momentum should be of the order of the 
infinitely large photon momentum. Having noted that, we again
apply Cutkosky's rules which yields the object of desire, the 
hadronic tensor. The temporal integral yields $k_0=q_0\pm\omega$ 
which then also fixes the 
spatial part of the loop momentum due to $\delta(k^2)$. To 
further exploit the Bjorken limit it is useful to introduce
the Fourier transformation of the quark wave--functions,
\be
\tilde{\Psi}_\alpha(\vec{p})=\int \frac{d^3x}{4\pi}\,
\Psi_\alpha(\vec{x})\, {\rm e}^{i\vec{x}\cdot\vec{p}}
\label{ftrans}
\ee
and make use of the fact that the eigenstates of the Dirac 
Hamiltonian carry definite parity. The latter property 
allows us to compensate spatial reflections by factors of 
$\beta$. With these preliminaries the spatial integrals
can be performed
\be
W_{\mu\nu}(q)&=&iM_N\frac{N_C}{2}(4\pi)^2\int \frac{d\omega}{2\pi}
\sum_\alpha \int \frac{d^3k}{(2\pi)^2}\frac{1}{2|\vec{k}|}
\label{wten0} \\ && \times
\Big\langle N\Big|\Bigg\{\Big[
\tilde{\Psi}_\alpha^\dagger(\vec{q}+\vec{k}){\cal Q}_A^2
\beta\gamma_\mu\kslash\gamma_\nu
\tilde{\Psi}_\alpha(\vec{q}+\vec{k})
\delta(|\vec{k}|-q_0-\omega)
\nonumber \\ && \hspace{2cm}
-\, \tilde{\Psi}_\alpha^\dagger(\vec{q}+\vec{k}){\cal Q}_A^2
\gamma_\nu\kslash\gamma_\mu\beta
\tilde{\Psi}_\alpha(\vec{q}+\vec{k})
\delta(|\vec{k}|-q_0+\omega)\Big]f_\alpha^+(\omega)\Big|_{\rm pole}
\nonumber \\ && \hspace{1.5cm}
+\,\Big[\tilde{\Psi}_\alpha^\dagger(\vec{q}+\vec{k}){\cal Q}_A^2
(\gamma_\mu\kslash\gamma_\nu)_5\beta
\tilde{\Psi}_\alpha(\vec{q}+\vec{k})
\delta(|\vec{k}|-q_0-\omega)
\nonumber \\ && \hspace{2cm}
-\, \tilde{\Psi}_\alpha^\dagger(\vec{q}+\vec{k}){\cal Q}_A^2
\beta(\gamma_\nu\kslash\gamma_\mu)_5
\tilde{\Psi}_\alpha(\vec{q}+\vec{k})
\delta(|\vec{k}|-q_0+\omega)\Big]f_\alpha^-(\omega)
\Big|_{\rm pole}\Big| N\Big\rangle\, .
\nonumber
\ee
The subscript `pole' indicates that the spectral integral
is restricted to those values of $\omega$ which cause
the respective denominators of the spectral functions
(\ref{sfct}) to vanish. It is interesting to remark that
those (unphysical) poles which explicitly contain the
cut--off, $\Lambda_i$, occur in positive and negative pairs. 
Hence there is no clear distinction between quark and anti--quark 
contributions. Note that the second term in (\ref{wten0})
not only differs from the first one in the spectral function 
but also with regard to the position of the Dirac matrix $\beta$. 

We fix the coordinate system by agreeing that the photon moves 
along the $z$--axis. This also introduces Bjorken's scaling variable,
\be
q_0=q_3-M_N\, x\, .
\label{bjx}
\ee
Furthermore we change the integration variable from 
$\vec{k}$ to $\vec{p}=\vec{q}+\vec{k}$. Taking into account 
the Jacobian for this transformation, which becomes feasible 
because only the regime $|\vec{p}|\approx0$ contributes to the 
integral \cite{Ja75,We96}, the hadronic tensor becomes
\be
W_{\mu\nu}(q)&=&i\frac{N_C}{4}
\int \frac{d\omega}{2\pi}\sum_\alpha \int d^3p\,
%\\&& \hspace{2.0cm}\times
\Big\langle N\Big|\Bigg\{\Big[
\tilde{\Psi}_\alpha^\dagger(\vec{p}){\cal Q}_A^2
\beta\gamma_\mu\nslash\gamma_\nu\tilde{\Psi}_\alpha(\vec{p})
f_\alpha^+(\omega)\Big|_{\rm pole}
\label{wten1}\\ && \hspace{3cm}
+\tilde{\Psi}_\alpha^\dagger(\vec{p}){\cal Q}_A^2
(\gamma_\mu\nslash\gamma_\nu)_5
\beta\tilde{\Psi}_\alpha(\vec{p})
f_\alpha^-(\omega)\Big|_{\rm pole}\Big]
\delta(p_3-M_Nx+\omega)
\nonumber \\ && \hspace{4cm}-\,\Big[
\tilde{\Psi}_\alpha^\dagger(\vec{p}){\cal Q}_A^2
\gamma_\nu\nslash\gamma_\mu\beta\tilde{\Psi}_\alpha(\vec{p})
f_\alpha^+(\omega)\Big|_{\rm pole}
\nonumber \\ && \hspace{3cm}
+\tilde{\Psi}_\alpha^\dagger(\vec{p}){\cal Q}_A^2
\beta(\gamma_\nu\nslash\gamma_\mu)_5
\tilde{\Psi}_\alpha(\vec{p})
f_\alpha^-(\omega)\Big|_{\rm pole}\Big]
\delta(p_3-M_Nx-\omega)
\Bigg\}\Big| N\Big\rangle\, .
\nonumber
\ee
Here we have introduced the light--cone vector 
${\rm n}^\mu=(1,0,0,1)^\mu$. Finally we adopt an exponential 
representation for the $\delta$--function which allows
us to return to a coordinate space representation 
\be
W_{\mu\nu}(q)&=&-iM_N\frac{N_C}{4}
\int \frac{d\omega}{2\pi}\sum_\alpha \int d^3 \xi
\int \frac{d\lambda}{2\pi}\, {\rm e}^{iM_Nx\lambda}
\label{wten} \hspace{6cm}~
\\* && \hspace{2.0cm}\times\Big\langle N\Big|
\Bigg\{\Big[\bar{\Psi}_\alpha({\vec\xi}){\cal Q}_A^2
\gamma_\mu\nslash\gamma_\nu\Psi_\alpha(\xipl)
{\rm e}^{-i\lambda\omega}
\nonumber \\ && \hspace{4cm}
-\bar{\Psi}_\alpha({\vec\xi}){\cal Q}_A^2
\gamma_\nu\nslash\gamma_\mu
\Psi_\alpha(\ximl)
{\rm e}^{i\lambda\omega}\Big]
f_\alpha^+(\omega)\Big|_{\rm pole}
\nonumber \\ && \hspace{3.0cm}
+\Big[\bar{\Psi}_\alpha({\vec\xi}){\cal Q}_A^2
(\gamma_\mu\nslash\gamma_\nu)_5
\Psi_\alpha(\ximl)
{\rm e}^{-i\lambda\omega}
\nonumber \\ && \hspace{4cm}
-\bar{\Psi}_\alpha({\vec\xi}){\cal Q}_A^2
(\gamma_\nu\nslash\gamma_\mu)_5
\Psi_\alpha(\xipl)
{\rm e}^{i\lambda\omega}\Big]f_\alpha^-(\omega)\Big|_{\rm pole}
\Bigg\}\Big| N\Big\rangle\, ,
\nonumber
\ee
which is similar to the decomposition into quark and 
anti--quark distributions.  Again, in the unregularized case, 
this simplifies enormously because $f_\alpha^-(\omega)=0$ while 
$f_\alpha^+(\omega)\big|_{\rm pole}=-
4\pi i\delta(\omega-\epsilon_\alpha)$. Apparently the 
hadronic tensor then indeed becomes a sum of quark and 
anti--quark distributions. However, in the Pauli--Villars 
regularized scheme, we have additional contributions from
quark and anti--quark distributions with dispersion relations
which also contain the cut--offs, $\Lambda_i$. Hence they 
differ from those dispersion relations na\"{\i}vely expected from 
the solutions of the Dirac equation (\ref{diagh}).

In the next step the hadronic tensor is contracted with 
appropriate projectors which in turn provides the structure functions.
In the Bjorken limit these projectors become quite simple
and are given in table \ref{tab_1}.
\begin{table}[htb]
\large
\renewcommand{\arraystretch}{1.3}
\caption{\label{tab_1}\sf 
Projection operators which extract the leading twist piece of
the nucleon structure functions from the hadronic tensor. It should 
be remarked that the projectors given in the spin independent
cases are not general but rather presume the contraction 
with $S_{\mu\nu\rho\sigma}$ which is defined before 
eq (\protect\ref{defsign}). The last 
row denotes the spin orientation of the nucleon.}
\begin{center}
\begin{tabular}{c|c|c|c}
$f_1$ & $f_2$ & $g_1$ & $g_T=g_1+g_2$ \\
\hline
$-\frac{1}{2}g^{\mu\nu}$&
$-xg^{\mu\nu}$&
$\frac{i}{2M_N}\epsilon^{\mu\nu\rho\sigma}
\frac{q_\rho p_\sigma}{q\cdot s}$ &
$\frac{-i}{2M_N}\epsilon^{\mu\nu\rho\sigma}
s_\rho p_\sigma$ \\
\hline
${{\rm spin}\atop{\rm independent}}$&
${{\rm spin}\atop{\rm independent}}$
& $\vec{s}\parallel\vec{q}$ &
$\vec{s}\perp\vec{q}$
\end{tabular}
\end{center}
\renewcommand{\arraystretch}{1.0}
\normalsize
\end{table}
Note that within the Bjorken limit the Callan--Gross
relation, $f_2(x)=2xf_1(x)$, is automatically fulfilled.
The unpolarized structure function $f_1(x)$ then becomes
\be
f_1(x)&=&-iM_N\frac{N_C}{2}\int \frac{d\omega}{2\pi}
\sum_\alpha \int d^3\xi \int \frac{d\lambda}{2\pi}\,
{\rm e}^{iM_Nx\lambda}
\left(\sum_{i=0}^2c_i\frac{\omega+\epsilon_\alpha}
{\omega^2-\epsilon_\alpha^2-\Lambda_i^2+i\epsilon}\right)_{\rm pole}
\nonumber\hspace{2cm}~ \\ &&\hspace{0.3cm}\times
\Big\langle N\Big|
\bar{\Psi}_\alpha(\vec{\xi}){\cal Q}^2_A\nslash
\Psi_\alpha(\xipl){\rm e}^{-i\omega\lambda}
-\bar{\Psi}_\alpha(\vec{\xi}){\cal Q}^2_A\nslash
\Psi_\alpha(\ximl)){\rm e}^{i\omega\lambda}
\Big| N\Big\rangle
\nonumber \\ 
&=&i\frac{5}{36}M_NN_C\int \frac{d\omega}{2\pi}
\sum_\alpha \int \frac{d\lambda}{2\pi}\,
{\rm e}^{iM_Nx\lambda}
\left(\sum_{i=0}^2c_i\frac{\omega+\epsilon_\alpha}
{\omega^2-\epsilon_\alpha^2-\Lambda_i^2+i\epsilon}\right)_{\rm pole}
\label{f1x} \\* &&\hspace{0.3cm}\times
\int d^3\xi\,\left\{ \Psi^\dagger_\alpha(\vec{\xi})
\left(1-\alpha_3\right)\Psi_\alpha(\xipl)
{\rm e}^{-i\omega\lambda}
-\Psi^\dagger_\alpha(\vec{\xi})
\left(1-\alpha_3\right)\Psi_\alpha(\ximl)
{\rm e}^{i\omega\lambda}\right\}
\nonumber
\ee
to leading order in $1/N_C$. Here we have employed parity invariance
as well as the symmetry under grand--spin rotations to show that 
there is no iso--vector piece\footnote{\parbox[t]{16.0cm}{Upon 
expanding in $\lambda$ the bilocal matrix element can be expressed 
as an infinite sum of local matrix elements. Due to parity 
invariance only matrix elements like 
$$
(\lambda \partial_3)^{2n}-\alpha_3(\lambda \partial_3)^{2m+1} 
$$
will be non--vanishing. As this operator is even under grand--spin 
reflections, the accompanying isospin operator must be so too.}}. 
This is perfectly consistent with the Gottfried sum rule \cite{Go67} 
vanishing at this order. Note that the spectral function 
in (\ref{f1x}) is fully regularized. This indicates that this piece 
is associated to the real part of the bosonized action. In 
appendix D we also calculate the sub--leading order contribution
to the unpolarized structure function $f_1(x)$, {\it cf.} eq 
(\ref{f1tot}) which expectedly is of iso--vector character. In 
contrast to the iso--scalar piece (\ref{f1x}) that iso--vector
contribution does not undergo regularization in the sense that
the explicit dependence on the cut--off $\Lambda_i$ cancels.
Of course, from our knowledge on the static property calculations
in the NJL model it was to be expected that in case the iso--scalar 
piece is regularized the iso--vector piece is not and vice versa 
\cite{Al96}. In particular, these findings show that the structure 
function which enters the Gottfried sum rule should not be 
regularized in contrast to previous studies. In ref \cite{Wa98,Po99} 
this structure function has been treated analogously to the one 
of neutrino nucleon scattering associated with the Adler sum rule. 
As discussed in appendix D the latter indeed undergoes regularization. 
This example clearly exhibits that obtaining the formal expressions
for structure functions from the defining action is unavoidable 
in cases when there is no relation to a static nucleon property.

Now we turn to the leading order in $1/N_C$ piece to the polarized 
structure functions\footnote{See ref \cite{Ja95} for an overview 
on spin dependent nucleon structure functions.}. They are 
obtained from the anti--symmetric piece of the hadronic tensor
\be
W_{\mu\nu}^{\rm A}\hspace{-0.2cm}&=&\hspace{-0.2cm}
-M_N\frac{N_C}{2}\epsilon_{\mu\rho\nu\sigma}{\rm n}^\rho
\int \frac{d\omega}{2\pi} \sum_\alpha \int d^3\xi 
\int \frac{d\lambda}{2\pi}\, {\rm e}^{iM_Nx\lambda}
\left(\sum_{i=0}^2c_i\frac{\omega+\epsilon_\alpha}
{\omega^2-\epsilon_\alpha^2-\Lambda_i^2+i\epsilon}\right)_{\rm pole}
\hspace{2cm}~
\nonumber \\ &&\hspace{0.5cm}\times \Big\langle N\Big|
\bar{\Psi}_\alpha(\vec{\xi}){\cal Q}^2_A\gamma^\sigma\gamma_5
\Psi_\alpha(\xipl)){\rm e}^{-i\omega\lambda}
+\bar{\Psi}_\alpha(\vec{\xi}){\cal Q}^2_A\gamma^\sigma\gamma_5
\Psi_\alpha(\ximl)
{\rm e}^{i\omega\lambda}\Big| N\Big\rangle\, .
\label{wmna}
\ee
Again, the spectral function is fully regularized as it originates
from $f_\alpha^+(\omega)-f_\alpha^-(-\omega)$. Here the prescription 
(\ref{defsign}) has a major impact. Without this specification the 
relative sign between the spectral functions  would have been positive 
resulting in the spectral function $(\omega+\epsilon_\alpha)/
(\omega^2-\epsilon_\alpha^2+i\epsilon)$. In that case 
$W_{\mu\nu}^{\rm A}$ would have to be associated with unregularized 
imaginary part of the action. As we will see in the following 
section, such a result is not compatible with the sum rules. The 
reason is that the leading order (in $1/N_C$) contributions to 
the axial charges stem from the regularized real part of the 
action. Using the relevant projection operator given in 
table~\ref{tab_1} we find for the longitudinal 
polarized structure function
\be
g_1(x)\hspace{-0.2cm}&=&\hspace{-0.2cm} -i\frac{M_NN_C}{36} 
\Big\langle N\Big| I_3 \Big| N\Big\rangle
\int \frac{d\omega}{2\pi} \sum_\alpha \int d^3\xi 
\int \frac{d\lambda}{2\pi}\, {\rm e}^{iM_Nx\lambda}
\left(\sum_{i=0}^2c_i\frac{\omega+\epsilon_\alpha}
{\omega^2-\epsilon_\alpha^2-\Lambda_i^2+i\epsilon}\right)_{\rm pole}
\hspace{1cm}~ \nonumber \\ && \hspace{-1cm}\times
\left[\Psi^\dagger_\alpha(\vec{\xi})\tau_3
\left(1-\alpha_3\right)\gamma_5
\Psi_\alpha(\xipl)
{\rm e}^{-i\omega\lambda}
+\Psi^\dagger_\alpha(\vec{\xi})\tau_3
\left(1-\alpha_3\right)\gamma_5
\Psi_\alpha(\ximl)
{\rm e}^{i\omega\lambda}\right]\, ,
\label{g1x}
\ee
where we have substituted the matrix element (\ref{collq4})
of the collective coordinates, $A$, sandwiched between nucleon states.
Similarly the transverse combination $g_T(x)=g_1(x)+g_2(x)$ 
is given by
\be
g_T(x)\hspace{-0.2cm}&=&\hspace{-0.2cm} -i\frac{M_NN_C}{36}
\Big\langle N\Big| I_3 \Big| N\Big\rangle
\int \frac{d\omega}{2\pi} \sum_\alpha \int d^3\xi
\int \frac{d\lambda}{2\pi}\, {\rm e}^{iM_Nx\lambda}
\left(\sum_{i=0}^2c_i\frac{\omega+\epsilon_\alpha}
{\omega^2-\epsilon_\alpha^2-\Lambda_i^2+i\epsilon}\right)_{\rm pole}
\nonumber \\ && \hspace{1cm}\times
\left[\Psi^\dagger_\alpha(\vec{\xi})\tau_1
\alpha_1\gamma_5 \Psi_\alpha(\xipl)
{\rm e}^{-i\omega\lambda}
+\Psi^\dagger_\alpha(\vec{\xi})\tau_1
\alpha_1\gamma_5
\Psi_\alpha(\ximl)
{\rm e}^{i\omega\lambda}\right]\, .
\label{gtx}
\ee
In obtaining eqs (\ref{g1x}) and (\ref{gtx}) we have repeatedly 
made use of both parity invariance as well as the grand--spin
reflection symmetry. As a result, the leading order pieces of 
the polarized structure functions are of iso--vector character.

We conclude this section by presenting an unpolarized 
`structure function' which is not directly extracted from the 
Compton amplitude but nevertheless is of interest in the parton 
model picture. Reversing the relative sign of the contributions 
of the backward moving intermediate quarks, yields 
\be 
\bar{f}_1(x)&=&-M_N\frac{N_C}{2}\sum_\alpha 
{\rm sign}(\epsilon_\alpha) \int d^3\xi
\int \frac{d\lambda}{2\pi}\, {\rm e}^{iM_Nx\lambda}
\hspace{7cm}~ \nonumber \\ &&\hspace{0cm}\times
\Big\langle N\Big|
\bar{\Psi}_\alpha(\vec{\xi}){\cal Q}^2_A\nslash
\Psi_\alpha(\xipl)
{\rm e}^{-i\epsilon_\alpha\lambda}
+\bar{\Psi}_\alpha(\vec{\xi}){\cal Q}^2_A\nslash
\Psi_\alpha(\ximl)
{\rm e}^{i\epsilon_\alpha\lambda}\Big| N\Big\rangle\, .
\label{f1bar}
\ee
This distribution could be thought of arising from 
coupling to an anti--hermitian flavor operator stemming 
from the exchange of a charged gauge boson. Another 
example for such a distribution emerges in neutrino--nucleon 
scattering which eventually leads to the Adler sum rule.
In the unregularized version, the two different parts 
could be thought of being attributed to quark and 
anti--quark distributions. Therefore
this `structure function' will eventually be related
to the baryon number distribution. In this piece the 
regularization has dropped out which made the extraction of 
the pole contribution simple. Using again the symmetries
of the eigenfunctions of the Dirac Hamiltonian this
`structure function' turns out to be pure iso--scalar
\be
\bar{f}_1(x)&=&-M_N\frac{N_C}{2}
\sum_\alpha {\rm sign}(\epsilon_\alpha)
\int d^3\xi \int \frac{d\lambda}{2\pi}\, 
{\rm e}^{iM_Nx\lambda}\hspace{7cm}~
\nonumber \\ && \hspace{0cm}\times
\left[\Psi^\dagger_\alpha(\vec{\xi})
\left(1-\alpha_3\right)
\Psi_\alpha(\xipl)
{\rm e}^{-i\epsilon_\alpha\lambda}
+\Psi^\dagger_\alpha(\vec{\xi})
\left(1-\alpha_3\right)
\Psi_\alpha(\ximl)
{\rm e}^{i\epsilon_\alpha\lambda}\right]\, ,
\label{f1bar1}
\ee
where we have also omitted the factor associated with the
charge matrix ${\cal Q}$.

\bigskip
\stepcounter{chapter}
\leftline{\Large\it 6. Sum rules}
\medskip

Sum rules relate moments of the structure functions to static 
properties of hadrons and a consistently formulated model is 
required to satisfy these sum rules. Static properties
are obtained by computing matrix elements of the symmetry 
currents. Here we want to reflect on the sum rules for the
nucleon structure functions, in particular with regard
to the prescription (\ref{defsign}). As above, we will
only consider the vacuum contribution because the one of
the explicitly occupied valence level is not subject to 
regularization. 

We recall that the symmetry currents are extracted from 
the terms linear in the external source contained in
\be
\A_{\Lambda,{\rm R}}^{(1,v)}&=&
-i\frac{N_C}{2}\sum_{i=0}^2c_i
{\rm Tr}\,\Bigg\{\left(-\bDp\bDp_5+\Lambda_i^2\right)^{-1}
\nonumber \\* &&\hspace{3.5cm}\times
\left[\vslash T_{\rm v}^a\bDp_5-\bDp\vslash T_{\rm v}^a
+\aslash\gamma_5T_{\rm a}^a\bDp_5
+\bDp\aslash\gamma_5T_{\rm a}^a\right]\Bigg\}
\label{sr1}
\ee
for the real part. Here $T_{\rm v,a}^a$ denote appropriate
flavor generators for the vector (v) and axial--vector (a)
currents. Similarly the imaginary part of the action
contributes
\be
\A_{\Lambda,{\rm I}}^{(1,v)}=
-i\frac{N_C}{2}
{\rm Tr}\,\left\{\left(-\bDp\bDp_5\right)^{-1}
\left[\vslash T_{\rm v}^a\bDp_5+\bDp\vslash T_{\rm v}^a
+\aslash\gamma_5T_{\rm a}^a\bDp_5-
\bDp\aslash\gamma_5T_{\rm a}^a\right]\right\}\, .
\label{sr2}
\ee
Again, we confine the present discussion to the leading 
order terms in the $1/N_C$ expansion. The sub--leading
order is relegated to appendix D. Reading off the coefficients 
of the source terms yields the charge operators
\be
Q_\mu^a&=&-i\frac{N_C}{2}D^{ab}\int 
\frac{d\omega}{2\pi}\sum_\alpha
\Bigg\{\langle\alpha|T_{\rm v}^b\gamma_\mu\beta|\alpha\rangle
f_\alpha^-(\omega)+
\langle\alpha|\beta T_{\rm v}^b\gamma_\mu|\alpha\rangle
f_\alpha^+(\omega) \Bigg\}
\label{vcharge}
\ee
and
\be
Q_\mu^{(5)a}&=&i\frac{N_C}{2}D^{ab}\int 
\frac{d\omega}{2\pi}\sum_\alpha
\Bigg\{\langle\alpha|T_{\rm a}^b\gamma_\mu\gamma_5\beta|\alpha\rangle
f_\alpha^-(\omega)+
\langle\alpha|\beta T_{\rm a}^b\gamma_\mu\gamma_5|\alpha\rangle
f_\alpha^+(\omega)\Bigg\}\,.
\label{acharge}
\ee
Here we have introduced the adjoint representation of
the collective coordinates for the soliton orientation, 
$D^{ab}=(1/2) {\rm tr}(A^\dagger\tau^a A\tau^b)$, 
which still has to be sandwiched 
between nucleon states, {\it cf.}~eq~(\ref{collq4}).
The similarity to the structure function expressions becomes
apparent in the reappearance of the spectral functions
(\ref{sfct}). Adopting symmetric contours for the Cauchy 
integrals immediately yields the sum over all poles contained
in the spectral functions $f_\alpha^{\pm}$ and the connection
to the hadronic tensor (\ref{wten0}) is evident.
Also note that in those matrix elements which are multiplied by
$f_\alpha^-$, the Dirac matrix $\beta$ and the adjoint state 
$\langle \alpha|$ do not combine to~$\bar{\Psi}_\alpha$.

The axial charges of the nucleon are obtained from the
spatial components, $Q_i^{(5)a}$. While the iso--scalar
part ($T_{\rm a}^0=\ID$) vanishes at this order of the 
$1/N_C$ expansion the iso--vector combination 
($T_{\rm a}^0=\tau^a/2$) results in the vacuum contribution
to the axial charge of the nucleon
\be
g_{\rm A}=\frac{N_C}{6}\Big
\langle N\Big|2I_3\Big|N\Big\rangle
\sum_{i=0}^2c_i\sum_\alpha
\frac{\epsilon_\alpha}{\sqrt{\epsilon_\alpha^2+\Lambda_i^2}}
\langle\alpha|\tau_3\alpha_3\gamma_5|\alpha\rangle\, .
\label{ga3}
\ee
As remarked above, this quantity originates from the real
part of the action as can be observed by the appearance
of the Pauli--Villars regulator. In order to verify the
sum rule for $g_{\rm A}$ the expression (\ref{ga3}) has to be
compared with the integral\footnote{The upper boundary
being infinity is subject to the non--locality of the
soliton. Eventually we have to boost the structure functions 
to the infinite momentum frame which properly projects
them the interval $x\in[0,1]$, {\it cf.} ref~\cite{Ga98}.} 
$\int_0^\infty dx g_1(x)$, which is computed in several 
steps. We recognize that the two terms in 
square brackets in eq (\ref{g1x}) are related by 
$\lambda\leftrightarrow -\lambda$ which allows us to 
extend the $x$--integration to minus infinity yielding
$(2\pi/M_N)\delta(\lambda)$ and whence a local matrix 
element. After performing the trivial $\lambda$--integration 
the odd powers of $\omega$ in the spectral function disappear
\be
\int_0^\infty g_1(x) &=&  i \frac{N_C}{36}
\langle N\Big|I_3\Big|N\Big\rangle
\int \frac{d\omega}{2\pi}\sum_\alpha 
\left(\sum_{i=0}^2c_i\frac{\epsilon_\alpha}
{\omega^2-\epsilon_\alpha^2-\Lambda_i^2+i\epsilon}\right)_{\rm pole}
\hspace{2cm}~\nonumber \\ &&\hspace{4cm}\times \int d^3\xi\,
\Psi^\dagger_\alpha(\vec{\xi})\tau_3\alpha_3\gamma_5
\Psi_\alpha(\vec{\xi})\,.
\label{g1xint}
\ee
The quark matrix element of $\tau_3\gamma_5$ has vanished because
of parity invariance. Also, the poles can easily be collected
\be
\left(\frac{1}
{\omega^2-\epsilon_\alpha^2-\Lambda^2+i\epsilon}\right)_{\rm pole}
&=&-\frac{i\pi}{\sqrt{\epsilon_\alpha^2+\Lambda^2}}\left[
\delta\left(\omega+\sqrt{\epsilon_\alpha^2+\Lambda^2}\right)
+\delta\left(\omega-\sqrt{\epsilon_\alpha^2+\Lambda^2}\right)
\right]\, .\qquad
\label{poles}
\ee
Hence the Bjorken sum rule
\be
\int_0^\infty dx \left(g_1^{\rm p}(x)-g_1^{\rm n}(x)\right)
=\frac{1}{6}g_{\rm A}
\label{bjsum}
\ee
is straightforwardly obtained after taking care of the isospin 
matrix elements of the nucleon. Using rotational invariance in 
grand--spin space in addition to the above explained calculational 
rules verifies the Burkhardt--Cottingham \cite{Bu70} sum rule
\be
\int_0^\infty dx\, g_2(x) = 0\, .
\label{bcsum}
\ee

Here we wish to re--emphasize that the prescription (\ref{defsign}) 
is crucial for deriving these results. Without it, the unregularized 
spectral function, $f^+(\omega)+f^-(-\omega)$, rather than
$f^+(\omega)-f^-(-\omega)$ would have appeared in (\ref{g1xint}). 
The former would not have resulted in the regularized form 
for $g_1(x)$ as required by the Bjorken sum rule. 

Although the treatment can be generalized to three flavors,
we are currently focusing on two flavor case. Hence
we require for the momentum sum rule 
\be
\frac{9}{5}\int_0^\infty dx \left(f_2^{\rm ep}+
f_2^{\rm en}\right)
=\frac{36}{5}\int_0^\infty dx\, x f_1(x)\, ,
\label{msr1}
\ee
according to the Callan--Gross relation and 
with $f_1(x)$ given in eq (\ref{f1x}). The factor $x$ can
be treated by differentiating with respect to $\lambda$. 
Furthermore the contribution from the backward propagating
quarks serves to complete the $x$--integral along the 
negative half--line. That is,
\be
\frac{36}{5}\int_0^\infty dx\, x f_1(x)&=&
-N_C\int \frac{d\omega}{2\pi}
\sum_\alpha \int \frac{d\lambda}{2\pi}\,
\int_{-\infty}^\infty dx\, {\rm e}^{iM_N x\lambda}\,
\left(\sum_{i=0}^2c_i\frac{\omega+\epsilon_\alpha}
{\omega^2-\epsilon_\alpha^2-\Lambda_i^2+i\epsilon}\right)_{\rm pole}
\nonumber \\* && \hspace{1cm}\times
\frac{\partial}{\partial \lambda}
\int d^3\xi\,\Psi^\dagger_\alpha(\vec{\xi})
\left(1-\alpha_3\right)\Psi_\alpha(\xipl)
{\rm e}^{-i\omega\lambda}
\label{msr2} \\ 
&=&-\frac{N_C}{M_N}
\int \frac{d\omega}{2\pi i}
\sum_\alpha \left(\sum_{i=0}^2c_i\frac{\omega+\epsilon_\alpha}
{\omega^2-\epsilon_\alpha^2-\Lambda_i^2+i\epsilon}\right)
\langle\alpha|\left(1-\alpha_3\right)
\left(i\partial_3 +\omega\right)|\alpha\rangle\, .
\nonumber
\ee
The subscript `pole' has been dropped because without
the oscillating function ${\rm e}^{\pm i\omega\lambda}$
the spectral integral directly sums up the pole terms.
The matrix elements of $i\partial_3$ and $\alpha_3\omega$
vanish due to parity invariance while the regularized 
sum of the matrix elements of $\alpha_3i\partial_3$ is zero 
for the field configuration which minimizes the soliton field 
energy\footnote{We recall that we contribution of the 
valence quark orbit is always carried along.}\cite{Di96}.
We are now left with
\be
\frac{36}{5}\int_0^\infty dx\, x f_1(x)=
-\frac{N_C}{M_N} \int \frac{d\omega}{2\pi i}
\sum_\alpha \left(\sum_{i=0}^2c_i\frac{\omega^2}
{\omega^2-\epsilon_\alpha^2-\Lambda_i^2+i\epsilon}\right)\, .
\label{msr3}
\ee
Up to the factor $1/M_N$ this can be shown to be the vacuum
contribution to the soliton energy (\ref{etot}) when using 
the regularization condition $\sum_{i=0}^2 c_i=0$, {\it cf.} 
eq (\ref{pvcond}). Hence we have verified the momentum
sum rule. In particular we see that the model saturates this
sum rule completely and there is no room for additional 
degrees of freedom, as {\it e.g.} gluons. In order to satisfy 
the momentum sum rule the iso--scalar content of $f_1$ 
apparently needs to be regularized. From the analogy to the 
computation of static properties we hence expect the iso--vector 
piece not to be regularized. This is, of course, perfectly 
consistent with our finding that the structure function entering 
the Gottfried sum rule,
$f_2^{\rm ep}-f_2^{\rm en}=2x(f_2^{\rm ep}-f_2^{\rm en})$
remains unregularized as discussed in  appendix D.

Above we have explicitly shown the necessity of the 
additional specification (\ref{defsign}) to render the 
sum rules of the polarized structure functions consistent
with the axial charges. On the formal level the above discussion 
is actually more transparent. When considering 
the Bjorken limit (\ref{simple6}) and (\ref{simple7}) we have 
already succeeded in expressing the two photon coupling in 
form of a linear coupling. What remains is the comparison of
the structure of the Dirac matrices in 
$\vslash\left(i\dslash\right)^{-1}\vslash$
and $(\vslash\left(i\dslash\right)^{-1}\vslash)_5$
with that in the above expressions. Note that in eqs 
(\ref{simple6}) and (\ref{simple7}) we have retained 
the full dependence of the Dirac operator on the pion fields.
In particular, for the soliton sector we have not yet carried out 
the expansion in the angular velocities (\ref{collq2}),(\ref{nuc4})
and (\ref{nuc4a}). According to eq (\ref{defsign}) these
contributions are given by\footnote{The additional factor `$i$' is 
contained in the projectors which are listed in table \ref{tab_1}.}
\be
(\vslash\left(i\dslash\right)^{-1}\vslash)_{(5)}
={\cal S}\cdot {\cal V} \pm {\cal E}\cdot {\cal A}\gamma_5\, .
\label{gendis1}
\ee
Then the sum of (\ref{simple6}) and (\ref{simple7}) can be 
expressed as (up to the overall factor $N_C/4i$)
\be
\A^{(2,v)}_\Lambda&=&
\sum_{i=0}^2c_i
{\rm Tr}\,\Bigg\{\left(-\bDp\bDp_5+\Lambda_i^2\right)^{-1}
\nonumber \\ &&\hspace{2cm}\times
\left[{\cal Q}^2\left({\cal S}\cdot {\cal V} + 
{\cal E}\cdot {\cal A}\gamma_5\right)\bDp_5
-\bDp\left({\cal S}\cdot {\cal V} - 
{\cal E}\cdot {\cal A}\gamma_5\right){\cal Q}^2\right]\Bigg\}
\nonumber \\ &&
+\,{\rm Tr}\,\left\{\left(-\bDp\bDp_5\right)^{-1}
\left[{\cal Q}^2\left({\cal S}\cdot {\cal V} +
{\cal E}\cdot {\cal A}\gamma_5\right)\bDp_5
+\bDp\left({\cal S}\cdot {\cal V} -
{\cal E}\cdot {\cal A}\gamma_5\right){\cal Q}^2\right]\right\}
\nonumber \\
&=&
\sum_{i=0}^2c_i
{\rm Tr}\,\Bigg\{\left(-\bDp\bDp_5+\Lambda_i^2\right)^{-1}
\nonumber \\ && \hspace{2cm} \times
\left[{\cal Q}^2{\cal S}\cdot {\cal V}\bDp_5
-\bDp{\cal S}\cdot {\cal V}{\cal Q}^2
+{\cal Q}^2 {\cal E}\cdot {\cal A}\gamma_5\bDp_5
+\bDp {\cal E}\cdot {\cal A}\gamma_5{\cal Q}^2\right]\Bigg\}
\nonumber \\ &&
+\,{\rm Tr}\,\Bigg\{\left(-\bDp\bDp_5\right)^{-1}
\label{gendis3}\\ && \hspace{2cm}  \times
\left[{\cal Q}^2{\cal S}\cdot {\cal V}\bDp_5
+\bDp{\cal S}\cdot {\cal V}{\cal Q}^2
+{\cal Q}^2 {\cal E}\cdot {\cal A}\gamma_5\bDp_5
-\bDp {\cal E}\cdot {\cal A}\gamma_5{\cal Q}^2\right]\Bigg\}\,.
\nonumber
\ee
This expression is exactly of the same structure as the sum of 
(\ref{sr1}) and (\ref{sr2}). It is this identical structure which
renders the sum rules fulfilled. In case we had not introduced the 
relative sign in eq (\ref{defsign}), which propagates through to eq 
(\ref{gendis1}), the last terms in the square brackets would have 
the opposite sign and would fail to match with the sum rules. When 
comparing eqs (\ref{sr1}) and (\ref{sr2}) to (\ref{gendis3}) we 
have not yet made explicit the `position' of the $\beta$--matrices
in $\bDp$ and $\bDp_5$ rather we have kept these operators
in full. Thus this relative sign is not a matter of shuffling 
the Dirac matrix $\beta$.

We complete this section by noting that the sum rule
for the `structure function' $\bar{f}_1(x)$ indeed
gives the baryon number of the polarized vacuum, {\it i.e.}
\be
\int_0^\infty dx \bar{f}_1(x)= 
-\frac{N_C}{2}\sum_\alpha {\rm sign}
\left(\epsilon_\alpha\right)\, .
\label{f1bar2}
\ee

\bigskip
\stepcounter{chapter}
\leftline{\Large\it 7. Conclusions}
\medskip

In order to establish a self consistent regularization scheme for 
the hadron structure functions in the bosonized Nambu--Jona--Lasinio model 
we have studied the absorptive part of the Compton amplitude 
in Bjorken limit.  Noting that the Compton amplitude is defined as the 
hadronic matrix element of a time--ordered correlation function of the 
symmetry currents, it is straightforwardly  obtained from 
the {\em regularized} NJL action functional.  In this respect, 
adopting the Compton amplitude as the preferred  starting point 
for the structure function calculations yields {\em self consistent} 
results.

One constraint on regularizing the quark determinant is to 
maintain the anomaly structure of the quark loop when coupled
to external sources. This is accomplished by regularizing only 
the $\gamma_5$ even piece (real part in Euclidean space) while 
leaving the conditionally convergent odd piece (imaginary part 
in Euclidean space) un--regularized. In a formal sense this leads 
to a model for scalar quarks which have gradient interactions
thereby making the computation of the Compton amplitude more 
involved. Fortunately it turns out that in the Bjorken limit,
which extracts the leading twist contribution of the structure
functions, substantial simplifications occur. The reason is that
the propagator, which carries the infinitely large momentum of the 
virtual photon in the quark loop, can be taken to be that of a free 
massless quark. Eventually this simplification also leads to the 
scaling laws of the structure functions. It is not apparent that 
these laws are maintained under regularization. In this respect we 
have seen that the Pauli--Villars regularization scheme for the 
bosonized NJL model is well suited because the cut--off's essentially 
act as masses which are additive to the momenta. For propagators 
which carry large external momenta the regularizing cut--off's can 
hence be ignored. Regularization schemes which exponentially suppress 
large momenta do not necessarily exhibit this feature. When studying 
the structure functions we have also recognized that treating the 
$\gamma_5$ even and odd pieces differently imposes further 
constraints on the regularization procedure in order to 
consistently formulate the sum rules. As this specification only 
effects the polarized structure functions its necessity has not been 
observed in previous calculations of the pion structure functions. 
Here it is worth noting that a dissimilar regularization of the 
$\gamma_5$ even and odd pieces requires a careful treatment in 
other cases as well. One example is the inclusion of
$\omega$ vector meson in the NJL model \cite{We95,Do96}.

Having made this specification the calculation of nucleon
structure functions from the Compton amplitude in the
NJL chiral soliton model turned out to be sensible. 
The sum rules for the fully regularized structure functions 
are straightforwardly verified because Cutkosky's rules, 
which define the absorptive part of the amplitude, extract 
exactly the same poles from the spectral integral as do the 
Cauchy integrals for the matrix elements associated with static 
nucleon properties. We have found that in general the structure 
function cannot be written as linear combinations of quark and 
anti--quark distributions; rather they acquire contributions 
from various poles in the quark propagators which are associated 
with the Pauli--Villars cut--off's as the latter play the role of 
quark masses. Only those combinations of structure functions which 
are related to the $\gamma_5$ odd piece and hence are not regularized 
turn out to be simple sums of quark and ant--quark distributions. 
When including cranking corrections to generate nucleon quantum numbers 
from the chiral soliton we have seen that only either the iso--scalar 
or the iso--vector part of the considered structure function undergoes 
regularization. Of course, this is expected from the computation of
nucleon charges in this model. Surprisingly, in the case of the 
structure function which enters the Gottfried sum rule we have 
seen that it is the iso--scalar rather than the iso--vector part 
which undergoes regularization. This has not been expected as this 
iso--vector piece is very similar to the structure function entering 
the Adler sum rule for (anti) neutrino nucleon scattering, which
indeed gets regularized.

Certainly this study represents the first step towards a full
self--consistent calculation which will in turn require
numerical analyses. We have already noted that in the NJL
model many of the nucleon properties are dominated by 
the corresponding contribution of the distinct valence level 
which does not undergo regularization.  Hence we do not expect 
drastic changes in comparison with previous calculations. 
Nevertheless, in light of the fact that the iso--vector 
component of the unpolarized structure function $f_1$ 
remains unregularized suggests a thorough re--evaluation of 
the Gottfried sum rule is necessary. In addition, in the case of 
structure functions which are subject to regularization, and hence 
are more difficult to access, one might eventually confine oneself 
to the evaluation of low moments being related to products of 
quark matrix elements of local operators.  As an example we 
will consider $d_2$, the second moment of the twist--3 piece of the 
transverse polarized structure function, $g_2(x)$
which currently is of experimental interest \cite{SLAC}.

\bigskip
\leftline{\Large\it Acknowledgments}
\medskip
We are grateful to H. Reinhardt for the discussions 
and perspective during the many phases of this project.
In addition H.W. would like to thank R. L. Jaffe for interesting 
discussions. L.G. wishes to thank K. A. Milton for support
on this project.

\appendix

\bigskip
\stepcounter{chapter}
\leftline{\Large\it Appendix A: 
\parbox[t]{12.5cm}{Pion structure function from the 
Bethe--Salpeter amplitude}}
\medskip

In ref \cite{Da95} and section~3 the pion structure functions
were obtained from analyzing the Compton amplitude. In this appendix 
we will show that these structure functions can directly be obtained 
from the pion Bethe--Salpeter amplitude. This is completely
consistent with the regularization and does not require to 
introduce any transverse cut--off. As we will see this can be done 
in terms of light-cone wave--functions handled in a proper way. To this 
end we consider the $\gamma_5$--even Pauli--Villars regularized effective 
action, up to second order in the pion field. This yields the correlation 
function in terms of the Bethe--Salpeter bound state amplitude $\chi(p,k)$ 
defined by 
\begin{equation}  
{\cal A}_{NJL}\Big|_{\pi\pi}= {1\over 2}  \int { d^4 p\over (2\pi)^4}
 \int { d^4 k \over (2\pi)^4 } 
{\vec{\tilde\pi}}(p) \cdot {\vec{\tilde\pi}}(-p)
(p^2-m_\pi^2)
\chi(p,k) \, . 
\end{equation}	 
According to eqs (\ref{specfct}) and (\ref{gpcoup}) the Bethe--Salpeter 
amplitude in the Pauli--Villars regularization scheme is 
\begin{equation}
\chi(p,k)
= 4N_c {\rm i}g^2 {d\over d p^2} \Big\{ p^2
\sum_i c_i 
{1   \over k^2 - m^2 - \Lambda_i^2 + {\rm i}\epsilon}\,
{1\over (k-p)^2 - m^2 - \Lambda_i^2 + {\rm i}\epsilon} \Big\} 
\Big|_{p^2=m_\pi^2 }
\, . 
\end{equation}
Let us introduce  light--cone (LC) variables
\begin{eqnarray}
k^+ = k^0 + k^3 \, , \quad k^- = k^0 - k^3  \, , \quad d^4 k =
{1\over 2} dk^+ dk^- d^2 k_{\perp} \, .
\end{eqnarray}
Employing the definition\footnote{We could equivalently  
define the equal time wave--function 
$ \Psi_\pi^2 ( \vec k ) = \int d k_0 \chi (p,k)\Big|_{p^2=m_\pi^2} $
and go to the infinite momentum frame, $ p^0 \to P + m_\pi^2 /(2P) $, 
$ p^3 \to P $ and $ p_\perp = 0_\perp $, $k^3 = x P $. }
\begin{equation}
\Psi_\pi^2 (x,k_\perp ) = \int d k^-  
\chi(p,k) \Big|_{k^+ = m_\pi x, p^2=m_\pi} 
\end{equation}
for the LC wave--function we find
\begin{eqnarray}
\Psi_\pi^2 (x,k_\perp) &=& {4 N_c {\rm i} g^2\over x(1-x) }
{d\over d m_\pi^2} \Bigg\{m_\pi^2 
\int {dk^-\over 2}   
{m_\pi x\over k^- m_\pi x -{k_{\perp}^2+m^2+\Lambda_i^2-{\rm i}\epsilon}}
\nonumber \\ &&\hspace{2cm} \times
{1\over  k^-m_\pi(x-1)-m_\pi^2(x-1) 
- {k_{\perp}^2 + m^2 + \Lambda_i^2 -{\rm i} }}
\Big\} \, . 
\end{eqnarray}
The poles in the complex $k^-$ plane
are located at 
\begin{eqnarray}
k^- = { m^2 + k_\perp^2 + \Lambda_i^2  -{\rm i} \epsilon \over m_\pi x} 
\quad {\rm and}\quad
k^- = m_\pi+ 
{ m^2 + k_\perp^2 + \Lambda_i^2  -{\rm i} \epsilon \over 
m_\pi (x-1)} \, . 
\end{eqnarray}
For $x>1$ or $x<0$ both poles are above and below the real axis
respectively, and hence the integral vanishes in either case. For
$ 0 < x < 1 $ the integral yields the LC wave--function
\begin{eqnarray}
\Psi_\pi^2 (x, k_\perp ) = {4N_c g^2 \over (2\pi)^2} \pi
{d\over d m_\pi^2} \left\{ m_\pi^2 
\sum_i c_i {1\over k_\perp^2 + m^2 + \Lambda_i^2 -m_\pi^2 x(1-x) -{\rm
i} \epsilon } \right\} 
\, . 
\end{eqnarray}
Notice that, due to the Pauli--Villars subtractions, we have the
asymptotic behavior in the transverse momentum $k_\perp$,
\begin{eqnarray}
\Psi_\pi (x, k_\perp )^2 \sim {4N_c g^2 \over (2\pi)^2} \pi
{\sum_i c_i \Lambda_i^4 \over k_\perp^6} \, . 
\end{eqnarray}
This guarantees the convergence of the $k_\perp$ integral without 
introducing a transverse cut--off. Thus, the Pauli--Villars regulators 
automatically provide a form of a transverse cut-off. Upon integrating 
the transverse momentum we get the structure function
\begin{eqnarray}
F(x) = \int {d^2 k_\perp \over (2\pi)^2 }\Big\{ 
\Big(\frac{1}{3}\Big)^2 \Psi^2_{\pi,d}(x,k_\perp) 
+\Big(\frac{2}{3}\Big)^2 \Psi^2_{\pi,u}(x,k_\perp) \Big\} \, , 
\end{eqnarray}
which coincides with the result found in ref \cite{Da95} and 
eq (\ref{disp6}) of section~3.
The light--cone interpretation has been pursued before in the unregularized 
case (see {\it e.g.} \cite{Fr94} and refs therein) and more recently 
\cite{He98,BH99} within a LC quantization. In these cases transverse 
cut--off's were introduced, {\it a posteriori}. As we have shown above 
this is not necessary. Also, the LC wave--function approach is a bit 
subtle in the present context due to the necessary regularization. 
For instance, the well known LC convolution formulas for the elastic 
pion form factor \cite{Cl80}, are only valid here in conjunction with
the regularization. That is, by writing the LC wave--function as a 
Pauli--Villars sum  
\begin{equation}
\Psi_\pi^2(x,k_\perp)=\sum_i c_i \Psi_{\pi,i}^2(x,k_\perp) \, , 
\end{equation} 
we have the regularized null-plane formula,
$ F(q^2)= \sum_i c_i F_i (q^2)$ for the elastic pion form factor.
This is {\it not} the same as computing $\Psi(x,k_\perp)$ and 
employing the convolution formula. 

\bigskip
\stepcounter{chapter}
\leftline{\Large\it Appendix B: 
\parbox[t]{12.5cm}{Separation into free--massless 
and full quark propagators}}
\medskip

In this appendix we reinforce the analysis presented in section~4. 
There we argued that in the Bjorken limit one of the two propagators 
which are involved in the calculation of the structure functions can be
reduced to the free massless one. Here we do this by means of a symmetric 
version of the Wigner transformation (WT) as defined in ref \cite{WT84}. 
Although this technique or similar ones are well known in other 
fields, it is perhaps not very familiar in the present context. 
Because of that and in order to set up the notation we will
review it briefly. The main advantage of using this
method lies in the possibility of a systematic evaluation of higher 
twist contributions to structure functions in the soliton model.

Using translational invariance and localized baryon states $|B\rangle$
with normalization $ \langle B|B\rangle =1 $, one has the convenient
formula for the Compton amplitude of virtual photons in the rest 
frame of the soliton target 
\begin{equation}
(-i)T_{\mu \nu} (p,q) = 2M_N
\int d^3 R \int d^4 \xi e^{ i q \cdot \xi }
\Big\langle B \, \Big|
T \Big\{ J_\mu (R+\frac{\xi}{2}) J_\nu (R-\frac{\xi}{2}) \Big\} \,
\Big| \, B \Big\rangle \, , 
\label{symcomten}
\end{equation}
where, as usual, $J_\mu (x)= \bar q (x) {\cal Q} \gamma_\mu q(x) $.
This equation is equivalent to eq (\ref{nuc1}) if translational 
invariance is used. The form (\ref{symcomten}), however, is slightly 
more convenient to prove the announced result. As discussed in 
section~5 the vacuum contribution can be obtained by a suitable 
functional differentiation of the fermion determinant with respect to the
external sources. In the unregularized case we get
\begin{equation}
T_{\mu\nu} (p,q) \Big|_{\rm vac}
= -M_N N_c \int d^3 R
\int d^4 \xi e^{ {\rm i} q \cdot \xi} {\rm tr} \Big(
\gamma_\mu {\cal Q} \langle R+\frac{\xi}{2} | {\bf S} | R -\frac{\xi}{2}
\rangle  \gamma_\nu {\cal Q} \langle R-\frac{\xi}{2} | {\bf S} | R
+\frac{\xi}{2} \rangle \Big)\, +\, {\rm cr.} \, , 
\end{equation}
where the propagator operator is defined as  $ {\bf S}:= ({\rm i}
{\bf D})^{-1}$, the trace is taken with respect to Dirac and 
flavor degrees of freedom while ``cr.'' denotes the crossed 
contribution $( q \to -q , \mu \to \nu )$, {\it i.e.} backward 
moving quarks. Schematically we have expressions of the form
\begin{equation}
\int d^4 \xi e^{ {\rm i} q \cdot \xi }
\langle x+\frac{\xi}{2} | A | x -\frac{\xi}{2} \rangle
\langle x-\frac{\xi}{2} | B | x +\frac{\xi}{2} \rangle \, , 
\end{equation}
{\it i.e.}, the Fourier transformation of a product of bilocal functions.
To deal with these expressions properly we use the following symmetric
definition of the WT as considered in ref \cite{WT84}. 
For any bilocal operator $A$, with matrix elements 
$ \langle x | A | x' \rangle $ let us define its symmetric Wigner
representation as
\begin{equation}
A(x,p) =
\int {\rm d}^4 \xi \,e^{i\xi p}\langle x+\frac{\xi}{2}|
A|x-\frac{\xi}{2}\rangle \, . 
\end{equation}
Using its inverse,
\begin{equation}
\langle x + \frac{\xi}{2}|A|x-\frac{\xi}{2} \rangle =
\int {d^4 p\over (2\pi)^4} e^{-{\rm i}
\xi \cdot p } A(x,p) \, , 
\end{equation}
we get
\begin{equation}
\int d^4 \xi e^{ {\rm i} q \cdot \xi }
\langle x+\frac{\xi}{2} | A | x -\frac{\xi}{2} \rangle
\langle x-\frac{\xi}{2} | B | x +\frac{\xi}{2} \rangle =
\int \frac{d^4 p}{(2\pi)^4}  A (x, p) B(x,p-q) \, . 
\label{fourier}
\end{equation}
Notice that we have arbitrarily chosen the operator $B$ to carry the
photon momentum.
The product of two operators satisfies the following expression for its
WT
\begin{eqnarray}
 (AB)(x,p) &=&
e^{\frac{\rm i}{2}(\partial_p^A\cdot\partial_x^B-\partial_p^B
\cdot\partial_x^A)}A(x,p)B(x,p) 
\label{wtrule}
\\ &=&
A(x-\frac{\rm i}{2}\partial_p,p+\frac{\rm i}{2}\partial_x) B(x,p) 
\nonumber \\ &=&
\sum_{n=0}^\infty \sum_{m=0}^\infty { i^n \over 2^n n!}{ (-i)^m \over
2^m m!}
 {\partial^{n+m}  A(x,p) \over \partial p_{\nu_1} \cdots \partial
 p_{\nu_n}
  \partial x_{\mu_1} \cdots \partial x_{\mu_m }}
 {\partial^{n+m}  B(x,p) \over \partial x^{\nu_1} \cdots \partial x^{\nu_n}
  \partial p^{\mu_1} \cdots \partial p^{\mu_m }} \nonumber \, .
\end{eqnarray}
In the limit of large momentum $p$ one clearly gets
\begin{equation}
(AB)(x,p) \to A(x-\frac{\rm i}{2}\partial_p ,p) B(x,p) \, . 
\end{equation}
The WT of the Dirac operators ${\bf D}$ and ${\bf D}_5$ are
\begin{equation}
{\rm i}{\bf D}(x,p)= \pslash - {\cal M}(x) 
\qquad {\rm and} \qquad
{\rm i}{\bf D}_5 (x,p)= -\pslash - {\cal M}_5 (x) \, , 
\end{equation}
respectively. Applying the product rule, eq.(\ref{wtrule}) to the 
operator identity ${\rm i} {\bf D} {\bf S} =1 $ we obtain 
\begin{equation}
{\rm i}{\bf D}(x-\frac{\rm i}{2} \partial_p \, , \,
 p+\frac{\rm i}{2}\partial_x \, ) {\bf S} (x,p)=1 \, .
\end{equation}
Hence the WT of the propagator is given by
\begin{equation}
{\bf S} (x,p)={1\over \pslash+\frac{\rm i}{2}
\dslash_x - {\cal M}(x-\frac{\rm i}{2}\partial_p \, )} \, . 
\end{equation}
This expression is suitable for a derivative expansion in the field
${\cal M}(x)$ as follows
\begin{equation}
{\bf S} (x,p)= {\bf S}_0 (x,p) + {\bf S}_1 (x,p) + {\bf S}_2 (x,p) +
\dots  \, , 
\end{equation}
where the subscript indicates the number of derivatives in 
${\cal M}(x)$\footnote{This is effectively a semiclassical expansion of 
the propagator as can be seen by reinserting $\hbar$ in the previous 
expressions
$ \partial_x \to \hbar \partial_x $ and $ \partial_p \to \hbar
\partial_p $.}.
The derivatives act to the right. Therefore the lowest orders
in this derivative expansions are given by
\begin{eqnarray}
{\bf S}_0 (x,p) &=& {1\over \pslash -{\cal M}(x)} \, , \\
{\bf S}_1 (x,p) &=& \frac{\rm i}{2}{\bf S}_0 (x,p) \Big(
\dslash_x - \partial_x {\cal M}(x)\cdot \partial_p
\Big) {\bf S}_0 (x,p) \, .
\end{eqnarray}
The proper way to evaluate first and higher order expressions 
is by commuting the derivatives to the right until they
annihilate the unit operator. For instance, in the first order of
derivatives we have
\begin{equation}
{\bf S}_1 (x,p) = \frac{\rm i}{2} {\bf S}_0 (x,p)
\Big( \gamma_\mu {\bf S}_0 (x,p) \partial^\mu {\cal M}(x) +
 \partial^\mu {\cal M}(x) {\bf S}_0 (x,p) \gamma_\mu \Big) {\bf S}_0
(x,p) \, . 
\end{equation}
As we see the number of zeroth order propagators increases due to the
coordinate and momentum derivatives. It actually is this expansion
which is indicated in figure~\ref{fig4_2}. Now it is obvious that in 
the limit of large momentum $p$ higher order terms in ${\bf S}_n (x,p)$ 
can be neglected as compared to the zeroth order one, ${\bf S}_0 (x,p)$. 
Finally the latter can be replaced by the free massless one, ${\bf S}_F
(x,p) $. In summary,
\begin{equation}
{\bf S} (x,p) \to {\bf S}_F
(x,p) = {1\over \pslash}  \, . 
\end{equation}
Notice that in the expression (\ref{fourier}) only one operator contains 
the momentum $q$. Therefore, we finally get in the limit of large momentum 
$q$,
\begin{eqnarray}
T_{\mu\nu} (p,q) |_{\rm sea}  = 
N_c \int d^3 R
\int {d^4 p \over (2\pi)^4}{\rm tr} \Big(
\gamma_\mu {\cal Q} {\bf S}( R , p ) \gamma_\nu {\cal Q} 
{\bf S} (R,p-q) \Big) + {\rm cr.}\\
 \to  N_c \int d^3 R
 \int d^4 \xi e^{ {\rm i} q \cdot \xi} {\rm tr} \Big(
\gamma_\mu {\cal Q} \langle R+\frac{\xi}{2} | {\bf S} | R -\frac{\xi}{2}
\rangle  \gamma_\nu {\cal Q} \langle -\frac{\xi}{2} | {\bf S}_F | 
\frac{\xi}{2} \rangle \Big) + {\rm cr.} \, , 
\end{eqnarray}
where we have already used that the free propagator is invariant under 
translations. Thus, in the Bjorken limit we may replace that Dirac 
operator through which the large photon momentum flows by the free 
one. Also note that in section~4 we have made this replacement
before taking the functional derivatives with respect to the 
vector sources. In the present context, this corresponds to 
summing over the two possibilities that either of the two 
propagators carries the large momentum.

In the regularized case the argument is a bit more involved but
the result is similar. The only point is to identify the operators
$A$ and $B$. Using eq (\ref{wtrule}), we get
\begin{eqnarray}
-{\bf D}{\bf D}_5 (x,p) &=& -\Big( \pslash
+ \frac{\rm i}{2}\dslash_x - {\cal M}(x-\frac{\rm
i}{2}\partial_p ) \, \Big) \Big( \pslash+{\cal M}_5 (x) \Big) \\
&=&
-( \pslash -{\cal M}(x) ) ( \pslash +{\cal M}_5 (x) )
+ {\rm i} \dslash_x {\cal M}_5 (x) \, ,
\end{eqnarray}
where in the second line we have used that $ {\cal M} \gamma_\mu =
\gamma_\mu {\cal M}_5 $. Several cases for $A$ and $B$ have to be
considered. From the asymptotic limit of the product rule one
deduces that the WT of a derivative operator of a given order 
is a polynomial in $p$ of the same order. In the limit of large
momentum this polynomial dominates over the potential piece. 
Due to this reason one has the simplification 
${\bf D}{\bf D}_5 + \Lambda_i^2 \to  {\bf D}{\bf D}_5$ in the large 
momentum limit. Furthermore,
\begin{equation}
 i{\bf D}_5 ( -{\bf D}{\bf D}_5 + \Lambda_i^2 )^{-1} 
\to (i{\bf D})^{-1} \qquad {\rm and} \qquad
( -{\bf D}{\bf D}_5 + \Lambda_i^2 )^{-1} i {\bf D}
\to (i{\bf D}_5)^{-1} \, .
\end{equation}
In summary, the simplification given by eq.(\ref{simple6}) is 
supported by this analysis. Moreover, it makes evident that the 
Pauli--Villars regularization leads to the scaling behavior. This 
is not apparent in other schemes which do not have the regulating 
cut--off in the polynomial piece.

\bigskip
\stepcounter{chapter}
\leftline{\Large\it Appendix C: Sum rules in the $D_5$ model}
\medskip
In this appendix we illustrate that the
unphysical nature of the $\bD_5$ model is reflected by the
sum rules for the polarized structure functions. This discussion
will take place on the formal level and whence we ignore 
regularization. According to eq (\ref{act0}) the $\bD_5$ model 
is defined as
\be
\A_5=-i N_C {\rm Tr}\, {\rm log}\, \left\{i\bD_5\right\}\, .
\label{defd5m}
\ee
The static properties are obtained from
\be
\A^{(1,v)}&=&
iN_C{\rm Tr}\,\left\{\left(i\bDp\right)^{-1}
\left[\vslash T^a_{\rm v} + 
\aslash\gamma_5 T^a_{\rm a}\right]\right\}\, ,
\nonumber \\
\A_5^{(1,v)}&=&
-iN_C{\rm Tr}\,\left\{\left(i\bDp_5\right)^{-1}
\left[\vslash T^a_{\rm v} - 
\aslash\gamma_5 T^a_{\rm a}\right]\right\}
\label{d5stat}
\ee
within the physical $\bD$ and unphysical $\bD_5$ models,
respectively. Omitting, for the time being, the collective
coordinates, the charges are given by
\be
Q^a_\mu=
\sum_\alpha{\rm sign}\left(\epsilon_\alpha\right)
\langle \alpha | \beta \gamma_\mu (\gamma_5) T^a_{\rm v,a}
|\alpha \rangle
\quad {\rm and} \quad
Q^a_{5,\mu}=
-\sum_\alpha{\rm sign}\left(\epsilon_\alpha\right)
\langle \alpha | \gamma_\mu (-\gamma_5) \beta T^a_{\rm v,a}
|\alpha \rangle \, .
\label{d5charge}
\ee
For convenience, we have put vector and axial--vector
generators together. To study the sum rules we require the
hadronic tensor, $W_{\mu\nu;5}$ within the $\bD_5$ model. In 
analogy to eq (\ref{simple8}), $W_{\mu\nu;5}$ will be generated by
\be
\A_5^{(2,v)}=
i\frac{N_C}{2}
{\rm Tr}\,\left\{\left(\bDp_5\right)^{-1}
\left[{\cal Q}^2\vslash\left(-\dslash\right)^{-1}
\vslash\right]\right\}\, .
\label{d5hb}
\ee
It should be recalled that $(-i\dslash)^{-1}$ represents the free 
massless quark propagator in the $\bD_5$ model, see eq (\ref{defd5}). 
As can directly be observed, the difference among the pole
contributions between the $\bD$ and $\bD_5$ models 
\be
\left(\frac{\pm\omega+\epsilon_\alpha}
{\omega^2-\epsilon_\alpha^2+i\epsilon}\right)_{\rm pole}
=2\pi i \left(\pm\omega+\epsilon_\alpha\right)
\delta\left(\omega^2-\epsilon_\alpha^2\right)
=2\pi i\, {\rm sign}(\epsilon_\alpha)
\label{d5sfct}
\ee
has no effect. For the same reason the relative overall signs did 
not change when going from eq (\ref{d5stat}) to eq (\ref{d5charge}).
Besides the overall sign the only difference between $\A_5^{(2,v)}$ 
and (\ref{simple8}) is therefore the position of the Dirac 
matrix~$\beta$ contained in $\bDp_5$,
\be
W_{\mu\nu;5}\hspace{-0.2cm}&=&\hspace{-0.2cm}
-M_NN_C\sum_\alpha {\rm sign}(\epsilon_\alpha)
\int dt\, \int d^3\xi_1\,\int d^3\xi_2\,
\int \frac{d^4k}{(2\pi)^4}\,
{\rm e}^{it(q^0+k^0)}\,
{\rm e}^{-i(\vec{\xi}_1-\vec{\xi}_2)\cdot(\vec{q}+\vec{k})}\,
\delta\left(k^2\right)
\hspace{1cm}~\nonumber \\ && \hspace{0.5cm}\times
\Big\langle N \Big|
\Bigg\{\Psi^\dagger_\alpha(\vec{\xi}_1)
{\cal Q}_A^2\gamma_\mu\kslash\gamma_\nu
\beta\Psi_\alpha(\vec{\xi}_2)\,
{\rm e}^{i\epsilon_\alpha t}
-\Psi^\dagger_\alpha(\vec{\xi}_2)
{\cal Q}_A^2\gamma_\nu\kslash\gamma_\mu
\beta\Psi_\alpha(\vec{\xi}_1)\,
{\rm e}^{-i\epsilon_\alpha t}\Bigg\}
\Big| N\Big\rangle\, .
\label{d5wmn}
\ee
In comparison with the expressions (\ref{d5charge}) this
particular position of $\beta$ actually is a desired result.
When reducing the product of Dirac matrices
\be
W_{\mu\nu;5}\hspace{-0.2cm}&=&\hspace{-0.2cm}
-M_NN_C\sum_\alpha {\rm sign}(\epsilon_\alpha)
\int dt\, \int d^3\xi_1\,\int d^3\xi_2\,
\int \frac{d^4k}{(2\pi)^4}\,
{\rm e}^{it(q^0+k^0)}\,
{\rm e}^{-i(\vec{\xi}_1-\vec{\xi}_2)\cdot(\vec{q}+\vec{k})}\,
\delta\left(k^2\right)k^\sigma
\hspace{0.5cm}~\nonumber \\ && \hspace{0.2cm}\times
\Big\langle N \Big|
\Bigg\{S_{\mu\rho\nu\sigma}\Bigg[
\Psi^\dagger_\alpha(\vec{\xi}_1){\cal Q}_A^2\gamma^\rho
\beta\Psi_\alpha(\vec{\xi}_2)\,{\rm e}^{i\epsilon_\alpha t}
-\Psi^\dagger_\alpha(\vec{\xi}_2)
{\cal Q}_A^2\gamma^\rho\beta\Psi_\alpha(\vec{\xi}_1)\,
{\rm e}^{-i\epsilon_\alpha t}\Bigg]
\nonumber \\ && \hspace{0.5cm}
-i\epsilon_{\mu\rho\nu\sigma}\Bigg[
\Psi^\dagger_\alpha(\vec{\xi}_1){\cal Q}_A^2\gamma^\rho\gamma_5
\beta\Psi_\alpha(\vec{\xi}_2)\,{\rm e}^{i\epsilon_\alpha t}
+\Psi^\dagger_\alpha(\vec{\xi}_2)
{\cal Q}_A^2\gamma^\rho\gamma_5\beta\Psi_\alpha(\vec{\xi}_1)\,
{\rm e}^{-i\epsilon_\alpha t}\Bigg]
\Bigg\} \Big| N\Big\rangle
\label{d5wmn1}
\ee
we recognize that in comparison with the hadronic tensor in 
the physical model (\ref{nuc5}) there appears 
\underline{no~new~relative} sign between vector and 
axial--vector pieces. However, such a sign would be demanded to 
comply with the sum rules of (\ref{d5charge}) which indeed has a 
\underline{relative} sign between the vector and axial--vector 
pieces. We conclude that a world defined by $\bD_5$ the Bjorken sum 
rule would apparently read
\be
\int dx \left(g_1^{\rm p}(x)-g_1^{\rm n}(x)\right)
=-\frac{1}{6}g_{\rm A}\, .
\label{ga5bj}
\ee
This, of course, would be a major inconsistency which 
we have resolved by the prescription~(\ref{defsign}).

\bigskip
\stepcounter{chapter}
\leftline{\Large\it Appendix D: Cranking corrections to
structure functions}
\medskip

As for the leading order in the $1/N_C$ expansion, we 
confine ourselves to the vacuum contribution because the
piece which is due to the explicit occupation of the valence 
level has already been discussed previously \cite{We96}.

The computation of the cranking corrections to the 
nucleon structure functions starts with expanding
the expressions (\ref{simple6}) and (\ref{simple7}) 
to linear order in the angular velocities 
$\vec{\Omega}$, which are defined in eq 
(\ref{collq2}). In doing so, we will repeatedly use
the relations (\ref{nuc4}) and (\ref{nuc4a}).

To linear order in $\vec{\Omega}$ we then find that part of the 
action which is quadratic in the vector sources $v_\mu$ to be
\be
\A^{(2)}&=&-i\frac{N_C}{4}\int \frac{d\omega}{2\pi} \sum_\alpha
\Bigg\{\langle \omega,\alpha|{\cal Q}^2_A\beta
\vslash\left(i\dslash\right)^{-1}\vslash
|\omega,\alpha\rangle\, f_\alpha^+(\omega)
\nonumber \\ && \hspace{3cm}
+\langle \omega,\alpha|{\cal Q}^2_A
\left[\vslash\left(i\dslash\right)^{-1}\vslash\right]_5\beta
|\omega,\alpha\rangle\, f_\alpha^-(\omega)
\nonumber \\ && \hspace{1cm}
+\frac{i}{2}
\langle \omega,\alpha|\beta
\left[\vslash\left(i\dslash\right)^{-1}\vslash
\,\hat{t}\,{\cal Q}^2_A\tauom
-\tauom\,{\cal Q}^2_A\,\hat{t}\,
\vslash\left(i\dslash\right)^{-1}\vslash\right]
|\omega,\alpha\rangle\, f_\alpha^+(\omega)
\nonumber \\ && \hspace{1cm}
+\frac{i}{2}
\langle \omega,\alpha|
\left[\vslash\left(i\dslash\right)^{-1}\vslash\,
\hat{t}\,{\cal Q}^2_A\tauom
-\tauom\,{\cal Q}^2_A\,\hat{t}\,
\vslash\left(i\dslash\right)^{-1}
\vslash\right]_5\beta
|\omega,\alpha\rangle\, f_\alpha^-(\omega)
\nonumber \\ && \hspace{1cm}
+\sum_\beta 
\langle\alpha|\tauom|\beta\rangle\,
\Bigg[\langle \omega,\beta|{\cal Q}^2_A\beta
\vslash\left(i\dslash\right)^{-1}\vslash
|\omega,\alpha\rangle\, g_{\alpha\beta}^+(\omega)
\nonumber \\ && \hspace{4cm}
+\langle \omega,\beta|{\cal Q}^2_A
\left[\vslash\left(i\dslash\right)^{-1}\vslash\right]_5\beta
|\omega,\alpha\rangle\, g_{\alpha\beta}^-(\omega)\Bigg]\Bigg\}\, .
\label{avv1}
\ee
In addition to the spectral functions for the linear 
matrix elements (\ref{sfct}) we have introduced analogous
objects for the bi--linear matrix elements
\be
g_{\alpha\beta}^\pm(\omega)=\sum_{i=0}^2c_i
\frac{(\omega\pm\epsilon_\alpha)
(\omega\pm\epsilon_\beta)+\Lambda_i^2}
{(\omega^2-\epsilon_\alpha^2-\Lambda_i^2+i\epsilon)
(\omega^2-\epsilon_\beta^2-\Lambda_i^2+i\epsilon)}
\pm
\frac{(\omega\pm\epsilon_\alpha)
(\omega\pm\epsilon_\beta)}
{(\omega^2-\epsilon_\alpha^2+i\epsilon)
(\omega^2-\epsilon_\beta^2+i\epsilon)}\,.
\label{blsfct}
\ee
The first two terms in eq (\ref{avv1}), which do not contain 
the angular velocity $\vec{\Omega}$, have already been discussed 
in section 5. The terms, which contain the time
coordinate operator $\hat{t}$, originate from the time 
dependence of the collective coordinates (\ref{nuc4a}).
Finally the bi--linear pieces stem from expanding 
$\bDp$ and $\bDp_5$ to linear order in $\vec{\Omega}$, 
{\it cf.} eq (\ref{defd5}). In eq (\ref{avv1}) the 
abbreviation ${\cal Q}_A$ refers to the collectively
rotated charge matrix at a distinct point in time. The same
is true for the angular velocity. To be specific
\be
{\cal Q}_A=A^\dagger(0){\cal Q}A(0)
\qquad {\rm and} \qquad
\frac{i}{2}\tauom=A^\dagger(t)\frac{d}{dt}A(t)\,\Bigg|_{t=0}\, ,
\label{time}
\ee 
as the time--dependence of the collective coordinates has
been treated according to (\ref{nuc4a}).

Before further manipulating (\ref{avv1}) it is 
illuminating to discuss the relation to the sum rules
at sub--leading order. The Adler sum rule for 
(anti) neutrino nucleon scattering involves the
isospin operator. The constant of proportionality
which relates this operator to the collectively
rotated\footnote{This rotation is inherited by the
iso--vector--vector piece in the flavor matrix which is 
analogous to ${\cal Q}_A$.} angular velocities is the 
moment of inertia, $\alpha^2$, {\it cf.} eq (\ref{collq3}). 
The latter is obtained by expanding the regularized action 
to quadratic order in $\vec{\Omega}$,
\be
\frac{\partial^2 \A}{\partial \Omega_l\partial\Omega_m}
\Bigg|_{\vec{\Omega}=0}\hspace{-0.5cm}&=&
\frac{i}{2}N_C\sum_{i=0}^2c_i\int \frac{d\omega}{2\pi}
\sum_{\alpha\beta}\langle\alpha|\tau_l|\beta\rangle
\langle\beta|\tau_m|\alpha\rangle\,
\frac{\omega^2+\epsilon_\alpha\epsilon_\beta+\Lambda_i^2}
{(\omega^2-\epsilon_\alpha^2-\Lambda_i^2+i\epsilon)
(\omega^2-\epsilon_\beta^2-\Lambda_i^2+i\epsilon)}
\nonumber \\
&=&\frac{i}{4}N_C\int \frac{d\omega}{2\pi}
\sum_{\alpha\beta}
\langle\alpha|\tau_l|\beta\rangle
\langle\beta|\tau_m|\alpha\rangle
\left[g_{\alpha\beta}^+(\omega)
+g_{\alpha\beta}^-(\omega)\right]\, .
\label{momin1}
\ee
In the second step we have used the property that the spectral 
integrals over those pieces which are odd in $\omega$ 
vanish. Although in practice this Cauchy integral could 
easily be performed, the obvious relation to the sum of 
the spectral functions, $g^+ +g^-$, shows that the Adler 
sum rule will be satisfied. The reason is that in this 
channel the integration over $x$ will lead to the replacement 
$\vslash\left(i\dslash\right)^{-1}\vslash\to\beta$ after 
assuming the Bjorken limit. Similarly one might want to 
consider the axial singlet charge which receives a 
non--vanishing contribution only at sub--leading order in
$1/N_C$. The generating expression is obtained from 
eqs (\ref{sr1}) and (\ref{sr2}) by putting $T^a_{\rm a}=1$ 
and $a^\mu=(0,0,0,1)$ while all vector quantities are set
to zero. This yields
\be
&&-iN_C\Omega_3\int\frac{d\omega}{2\pi} \sum_{\alpha\beta}
\langle\alpha|\tau_3|\beta\rangle
\langle\beta|\alpha_3\gamma_5|\alpha\rangle\,
\frac{\omega^2+\epsilon_\alpha\epsilon_\beta}
{(\omega^2-\epsilon_\alpha^2+i\epsilon)
(\omega^2-\epsilon_\beta^2+i\epsilon)}
\nonumber \\ &&\hspace{3cm}
=iN_C\Omega_3\int\frac{d\omega}{2\pi} \sum_{\alpha\beta}
\langle\alpha|\tau_3|\beta\rangle
\langle\beta|\alpha_3\gamma_5|\alpha\rangle
\left[g_{\alpha\beta}^+(\omega)
-g_{\alpha\beta}^-(\omega)\right]\, .
\label{axsing}
\ee
As was to be expected the dependence on the cut--off,
$\Lambda_i$ dropped out because this quantity is associated
to the imaginary part of the action. However, it is 
remarkable that the difference $g^+-g^-$ of the 
spectral functions appeared rather than their sum. 
On the level of the hadronic tensor a difference between 
the last and next--to--last terms in (\ref{avv1}) can only 
be gained by the prescription (\ref{defsign}) since
the generator for the axial singlet charge 
($\gamma_3\gamma_5$) commutes with the Dirac
matrix $\beta$. Of course, in view of the previous
discussions this result was to be expected.

\begin{figure}[hbt]
\caption{\label{figD_1}\sf Summing the poles of 
the bi--local cranking corrections. The angular velocity
$\vec{\Omega}$ acts as a perturbation to the
quark propagator in the soliton background.}
~
\vskip0.2cm
\centerline{
\epsfig{figure=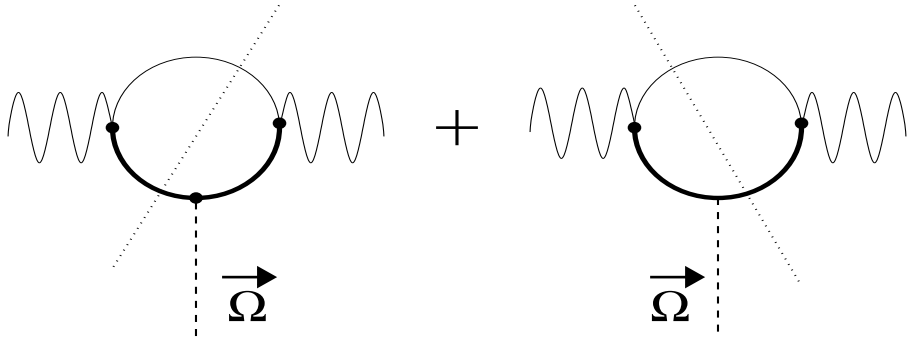,height=4.5cm,width=12cm}}
\end{figure}

We now continue to compute the hadronic tensor by varying 
(\ref{avv1}) with respect to the vector sources $v^\mu(\xi_1)$
and $v^\nu(\xi_2)$. At first sight, this would yield 
contributions from the forward and backward moving quarks 
for each single term in (\ref{avv1}). However, this is not the 
case for those terms which contain the time--coordinate operator 
$\hat{t}$ because it may act on a state like 
$|\vec{\xi},t=0\rangle$ which in turn gives zero contribution. 
In what follows we will concentrate on the treatment of that 
operator. At most it will contribute a factor linear in the 
time--coordinate which is integrated over when 
calculation the Compton tensor (\ref{Comp1}). For 
that calculation the coordinate $\xi=t$ can be 
expressed as a derivative with respect to $q^0$. In
turn this derivative can be re--written as a derivative
with respect to Bjorken $x$. In the nucleon rest frame 
we have $\partial x/\partial q^0=-1/2M_N$. Once 
this is established, the calculation of the 
hadronic tensor proceeds as in section 5 and
yields an expression analogous to eq (\ref{wten}).
Differentiating with respect to $x$ then simply
gives a factor $\lambda M_N$. In its full glory 
the hadronic tensor finally reads 
\be
W_{\mu\nu}&\bjlim&-iM_N\frac{N_C}{4}
\int \frac{d\omega}{2\pi}\sum_\alpha \int d^3 \xi
\int \frac{d\lambda}{2\pi}\, {\rm e}^{iM_Nx\lambda}\,
\Big\langle N\Big|
\label{wt1nc} \\ &&\hspace{-2cm}
\Bigg\{\Big[\bar{\Psi}_\alpha({\vec\xi}){\cal Q}_A^2
\gamma_\mu\nslash\gamma_\nu\Psi_\alpha(\xipl)
{\rm e}^{-i\lambda\omega}
-\bar{\Psi}_\alpha({\vec\xi}){\cal Q}_A^2
\gamma_\nu\nslash\gamma_\mu
\Psi_\alpha(\ximl)
{\rm e}^{i\lambda\omega}\Big]
f_\alpha^+(\omega)\Big|_{\rm p}
\nonumber  \\ &&\hspace{-2cm}
+\Big[\bar{\Psi}_\alpha({\vec\xi}){\cal Q}_A^2
(\gamma_\mu\nslash\gamma_\nu)_5
\Psi_\alpha(\ximl)
{\rm e}^{-i\lambda\omega}
-\bar{\Psi}_\alpha({\vec\xi}){\cal Q}_A^2
(\gamma_\nu\nslash\gamma_\mu)_5
\Psi_\alpha(\xipl)
{\rm e}^{i\lambda\omega}\Big]
f_\alpha^-(\omega)\Big|_{\rm p}
\nonumber \\ &&\hspace{-2.1cm}
+\frac{i\lambda}{4}\Big[\bar{\Psi}_\alpha({\vec\xi})
\tauom {\cal Q}_A^2 \gamma_\mu\nslash\gamma_\nu
\Psi_\alpha(\xipl) {\rm e}^{-i\lambda\omega}
+\bar{\Psi}_\alpha({\vec\xi})
{\cal Q}_A^2\tauom\gamma_\nu\nslash\gamma_\mu
\Psi_\alpha(\ximl) {\rm e}^{i\lambda\omega}\Big]
f_\alpha^+(\omega)\Big|_{\rm p}
\nonumber  \\ &&\hspace{-2.1cm}
+\frac{i\lambda}{4}\Big[\bar{\Psi}_\alpha({\vec\xi})
\tauom {\cal Q}_A^2(\gamma_\mu\nslash\gamma_\nu)_5
\Psi_\alpha(\ximl) {\rm e}^{-i\lambda\omega}
+\bar{\Psi}_\alpha({\vec\xi}){\cal Q}_A^2\tauom
(\gamma_\nu\nslash\gamma_\mu)_5
\Psi_\alpha(\xipl)
{\rm e}^{i\lambda\omega}\Big]
f_\alpha^-(\omega)\Big|_{\rm p}
\nonumber  \\ &&\hspace{-2.0cm}
+\sum_\beta\langle\alpha|\tauom|\beta\rangle 
\Bigg(\Big[\bar{\Psi}_\beta({\vec\xi}){\cal Q}_A^2
\gamma_\mu\nslash\gamma_\nu\Psi_\alpha(\xipl)
{\rm e}^{-i\lambda\omega}
-\bar{\Psi}_\beta({\vec\xi}){\cal Q}_A^2
\gamma_\nu\nslash\gamma_\mu
\Psi_\alpha(\ximl)
{\rm e}^{i\lambda\omega}\Big]
g_{\alpha\beta}^+(\omega)\Big|_{\rm p}
\nonumber  \\ &&\hspace{-1.2cm}
+\Big[\bar{\Psi}_\beta({\vec\xi}){\cal Q}_A^2
(\gamma_\mu\nslash\gamma_\nu)_5
\Psi_\alpha(\ximl)
{\rm e}^{-i\lambda\omega}
-\bar{\Psi}_\beta({\vec\xi}){\cal Q}_A^2
(\gamma_\nu\nslash\gamma_\mu)_5
\Psi_\alpha(\xipl)
{\rm e}^{i\lambda\omega}\Big]
g_{\alpha\beta}^-(\omega)\Big|_{\rm p}
\Bigg)\Bigg\}\Big| N \Big\rangle\, .
\nonumber
\ee
Here have made use of the fact that $\tauom$ does not 
change the parity of single quark states. This implies 
that in the double sum the states $|\alpha\rangle$ 
and $|\beta\rangle$ carry the same parity quantum 
number. In turn this allowed us to simply compensate
factors of the Dirac matrix $\beta$ by spatial 
reflections. In the double sum the pole contribution is 
obtained by summing over all possible frequencies 
which lead to singularities in the spectral functions 
(\ref{blsfct}). This summation is indicated in 
figure~\ref{figD_1}. 

At this point it is worth to briefly comment on the possible 
occurance of double poles, {\it i.e.} 
$\epsilon^2_\alpha=\epsilon^2_\beta+\Lambda_i^2$
and $\epsilon^2_\alpha=\epsilon^2_\beta$ in the regularized and 
un--regularized cases, respectively. According to Cutkosky's rules 
they contribute derivatives of $\delta$--functions to the absorptive
part of the amplitude: $-2\pi i \partial\, \delta
\left(\omega^2-\epsilon_\alpha^2\right)/\partial\omega^2$.
This is directly linked to the double poles which correspondingly
appear in the spectral integrals for the charges (\ref{axsing}) 
and the moment of inertia (\ref{momin1}). They can as well be
expressed as derivatives of single poles. In this way the
sum rules are ensured to be satisfied.

{\it A priori} the product $\tauom{\cal Q}^2_A$ is 
plagued by ordering ambiguities when imposing the 
quantization rules (\ref{collq3}) for the collective 
coordinates. To guarantee hermiticity we assume the
symmetric ordering. That is
\be
\frac{1}{2}\left\{J_i,{\cal Q}^2_A\right\}
=\frac{1}{18}\left(5J_i-I_3\tau_i\right)\, ,
\label{symord}
\ee
where we have also used that the matrix elements are evaluated 
in the subspace of nucleon states. This allowed us to employ 
eq (\ref{collq4}). In particular, the symmetric ordering
permitted us to identify $\tauom{\cal Q}^2_A={\cal Q}^2_A\tauom$.
For notational brevity we will maintain that expression in what
follows, keeping in mind that it should be substituted by 
(\ref{symord}). Contracting the hadronic tensor (\ref{wt1nc})
with the appropriate projector from table~\ref{tab_1} 
gives the unpolarized structure function to sub--leading
order in $1/N_C$, see also eq (\ref{f1x}),
\be
f_1(x)&=&-iM_N\frac{N_C}{2}\int \frac{d\omega}{2\pi}
\sum_\alpha \int d^3\xi \int \frac{d\lambda}{2\pi}\,
{\rm e}^{iM_Nx\lambda}\, \Big\langle N\Big|
\label{f1Nc}\hspace{2cm}~ \\ &&\hspace{-1.3cm}\times
\Bigg\{\left[\bar{\Psi}_\alpha(\vec{\xi}){\cal Q}^2_A\nslash
\Psi_\alpha(\xipl){\rm e}^{-i\omega\lambda}
-\bar{\Psi}_\alpha(\vec{\xi}){\cal Q}^2_A\nslash
\Psi_\alpha(\ximl)){\rm e}^{i\omega\lambda}\right]
\left(\sum_{i=0}^2c_i\frac{\omega+\epsilon_\alpha}
{\omega^2-\epsilon_\alpha^2-\Lambda_i^2+i\epsilon}\right)_{\rm p}
\nonumber \\ &&\hspace{-1.2cm}
+\frac{i\lambda}{4}\left[
\bar{\Psi}_\alpha(\vec{\xi})\tauom{\cal Q}^2_A\nslash
\Psi_\alpha(\xipl){\rm e}^{-i\omega\lambda}
+\bar{\Psi}_\alpha(\vec{\xi})\tauom{\cal Q}^2_A\nslash
\Psi_\alpha(\ximl)){\rm e}^{i\omega\lambda}\right]
\left(\frac{\omega+\epsilon_\alpha}
{\omega^2-\epsilon_\alpha^2+i\epsilon}\right)_{\rm p}
\nonumber \\ &&\hspace{-1.2cm}
+\sum_\beta\left[\bar{\Psi}_\beta(\vec{\xi})
{\cal Q}^2_A\nslash\Psi_\alpha(\xipl){\rm e}^{-i\omega\lambda}
-\bar{\Psi}_\beta(\vec{\xi}){\cal Q}^2_A\nslash
\Psi_\alpha(\ximl)){\rm e}^{i\omega\lambda}\right]
\nonumber \\ &&\hspace{1.5cm}\times
\langle\alpha|\tauom|\beta\rangle
\left(\frac{(\omega+\epsilon_\alpha)
(\omega+\epsilon_\beta)}
{(\omega^2-\epsilon_\alpha^2+i\epsilon)
(\omega^2-\epsilon_\beta^2+i\epsilon)}\right)_{\rm p}
\Bigg\}\,\Big| N\Big\rangle\, .
\nonumber
\ee
We observe that in the sub--leading contributions the
dependence on the cut-off, $\Lambda_i$ has disappeared
as the remaining spectral functions are associated with
\be
f_\alpha^+(\omega)+f_\alpha^-(-\omega)&=&
2\frac{\omega+\epsilon_\alpha}
{\omega^2-\epsilon_\alpha^2+i\epsilon}\, ,
\label{fgsum1} \\ 
g_{\alpha\beta}^+(\omega)-g_{\alpha\beta}^-(-\omega)&=&
2\frac{(\omega+\epsilon_\alpha)
(\omega+\epsilon_\beta)}
{(\omega^2-\epsilon_\alpha^2+i\epsilon)
(\omega^2-\epsilon_\beta^2+i\epsilon)}\,.
\ee
Actually this finding shows that previous computations
\cite{Wa98,Po99} regularizing the vacuum contribution to the 
Gottfried sum rule are not consistent with the present analysis 
of the Compton amplitude. This has to be contrasted with the 
iso--vector part of the structure function for the neutrino nucleon 
scattering. The only significant difference is the opposite sign for 
the terms involving the crossed ordering ($\nu\mu$), {\it i.e.} 
backward propagating quarks. The reason is that neutrino scattering 
involves the exchange of a charged gauge boson. Hence the iso--vector 
projection switches sign when going from the product of 
currents $J_\mu(\xi)J_\nu^\dagger(0)$ to the hermitian
conjugate $J_\nu(0)J_\mu^\dagger(\xi)$. Having noted that, 
we find the spectral function to be
\be
g_{\alpha\beta}^+(\omega)+g_{\alpha\beta}^-(-\omega)&=&
2\sum_{i=0}^2 c_i
\frac{(\omega+\epsilon_\alpha)
(\omega+\epsilon_\beta)}
{(\omega^2-\epsilon_\alpha^2-\Lambda_i^2+i\epsilon)
(\omega^2-\epsilon_\beta^2-\Lambda_i^2+i\epsilon)}
\label{fgsum2}
\ee
and hence the Adler sum rule to be satisfied, {\it cf.} eq 
(\ref{momin1}). 

We can now calculate the appropriate matrix elements 
by again using the grand--spin and parity properties 
of the single quark states,
\be
f_1(x)&=&{\rm eq~(\protect\ref{f1x})} +
i\, I_3\, \frac{M_NN_C}{36\alpha^2}\int \frac{d\omega}{2\pi}
\sum_\alpha \int \frac{d\lambda}{2\pi}\,
{\rm e}^{iM_Nx\lambda}\int d^3 \xi
\label{f1tot}\\* &&\hspace{-2cm} \times \Bigg\{
\frac{3i\lambda}{4}
\left[\Psi^\dagger_\alpha(\vec{\xi})
\left(\hspace{-0.5mm}1-\hspace{-0.5mm}\alpha_3\right)
\Psi_\alpha(\xipl) {\rm e}^{-i\omega\lambda}
+\Psi^\dagger_\alpha(\vec{\xi})
\left(\hspace{-0.5mm}1-\hspace{-0.5mm}\alpha_3\right)
\Psi_\alpha(\ximl){\rm e}^{i\omega\lambda}\right]
\left(\frac{\omega+\epsilon_\alpha}
{\omega^2-\epsilon_\alpha^2+i\epsilon}\right)_{\rm p}
\nonumber \\ &&\hspace{-1.5cm} 
+\sum_\beta\left(\frac{(\omega+\epsilon_\alpha)
(\omega+\epsilon_\beta)}
{(\omega^2-\epsilon_\alpha^2+i\epsilon)
(\omega^2-\epsilon_\beta^2+i\epsilon)}\right)_{\rm p}
\nonumber \\ &&\hspace{-1.0cm} \times
\Bigg(4\langle \alpha|\tau_1|\beta\rangle
\left[\Psi^\dagger_\alpha(\vec{\xi})\tau_1
\left(\hspace{-0.5mm}1-\hspace{-0.5mm}\alpha_3\right)
\Psi_\alpha(\xipl) {\rm e}^{-i\omega\lambda}
-\Psi^\dagger_\alpha(\vec{\xi})\tau_1
\left(\hspace{-0.5mm}1-\hspace{-0.5mm}\alpha_3\right)
\Psi_\alpha(\ximl){\rm e}^{i\omega\lambda}\right]
\nonumber \\ &&\hspace{-0.2cm}
-\langle \alpha|\tau_3|\beta\rangle
\left[\Psi^\dagger_\alpha(\vec{\xi})\tau_3
\left(\hspace{-0.5mm}1-\hspace{-0.5mm}\alpha_3\right)
\Psi_\alpha(\xipl) {\rm e}^{-i\omega\lambda}
-\Psi^\dagger_\alpha(\vec{\xi})\tau_3
\left(\hspace{-0.5mm}1-\hspace{-0.5mm}\alpha_3\right)
\Psi_\alpha(\ximl){\rm e}^{i\omega\lambda}\right]
\Bigg)\Bigg\}\, .
\nonumber 
\ee
The appearance of two terms in the cranking piece
reflects the fact that only an axial symmetry is left 
after selecting a direction for the photon momentum.
Also note that (\ref{f1tot}) does not contain the
nucleon spin operator. Of course, this is expected
for an unpolarized structure function. 

The anti--symmetric piece of the hadronic tensor reads
\be
W^{\rm A}_{\mu\nu}&=&-M_N\frac{N_C}{2}
\epsilon_{\mu\nu\rho\sigma} {\rm n}^\rho\,
\int \frac{d\omega}{2\pi}
\sum_\alpha \int d^3\xi \int \frac{d\lambda}{2\pi}\,
{\rm e}^{iM_Nx\lambda}\, \Big\langle N\Big|
\label{Wa1Nc}\hspace{2cm}~ \\ &&\hspace{-1.3cm}\times
\Bigg\{\Big[\bar{\Psi}_\alpha(\vec{\xi})
{\cal Q}^2_A\gamma^\sigma\gamma_5
\Psi_\alpha(\xipl){\rm e}^{-i\omega\lambda}
\nonumber \\ &&\hspace{3.0cm}
+\bar{\Psi}_\alpha(\vec{\xi})
{\cal Q}^2_A\gamma^\sigma\gamma_5
\Psi_\alpha(\ximl)){\rm e}^{i\omega\lambda}\Big]
\left(\sum_{i=0}^2c_i\frac{\omega+\epsilon_\alpha}
{\omega^2-\epsilon_\alpha^2-\Lambda_i^2+i\epsilon}\right)_{\rm p}
\nonumber \\ &&\hspace{-1.2cm}
+\frac{i\lambda}{4}\Big[
\bar{\Psi}_\alpha(\vec{\xi})\tauom
{\cal Q}^2_A\gamma^\sigma\gamma_5
\Psi_\alpha(\xipl){\rm e}^{-i\omega\lambda}
\nonumber \\ &&\hspace{3.0cm}
-\bar{\Psi}_\alpha(\vec{\xi})\tauom
{\cal Q}^2_A\gamma^\sigma\gamma_5
\Psi_\alpha(\ximl)){\rm e}^{i\omega\lambda}\Big]
\left(\frac{\omega+\epsilon_\alpha}
{\omega^2-\epsilon_\alpha^2+i\epsilon}\right)_{\rm p}
\nonumber \\ &&\hspace{-1.2cm}
+\sum_\beta\left[\bar{\Psi}_\beta(\vec{\xi})
{\cal Q}^2_A\gamma^\sigma\gamma_5
\Psi_\alpha(\xipl){\rm e}^{-i\omega\lambda}
+\bar{\Psi}_\beta(\vec{\xi})
{\cal Q}^2_A\gamma^\sigma\gamma_5
\Psi_\alpha(\ximl)){\rm e}^{i\omega\lambda}\right]
\nonumber \\ &&\hspace{3.0cm}\times
\langle\alpha|\tauom|\beta\rangle
\left(\frac{(\omega+\epsilon_\alpha)
(\omega+\epsilon_\beta)}
{(\omega^2-\epsilon_\alpha^2+i\epsilon)
(\omega^2-\epsilon_\beta^2+i\epsilon)}\right)_{\rm p}
\Bigg\}\,\Big| N\Big\rangle\, .
\nonumber
\ee
We note that after contracting the cranking corrections with 
the pertinent projectors from table~\ref{tab_1}, the parity 
properties of the single quark states enforce the combination 
of the Dirac matrices and the gradient expansion in $\lambda$
to be odd under grand--spin reflection ($\gamma_5$ is even 
under this operation). Hence the isospin part must be odd
as well. Therefore the cranking corrections to the polarized
structure functions are iso--singlets in the space of the 
collective coordinates. To be specific,
\be
g_1(x)&=&{\rm eq~(\protect\ref{g1x})} +
\frac{5i}{6\alpha^2}M_NN_C\int \frac{d\omega}{2\pi}
\sum_\alpha \int \frac{d\lambda}{2\pi}\,
{\rm e}^{iM_Nx\lambda}\int d^3 \xi
\label{g1tot} \\ &&\hspace{0cm} \times \Bigg\{
\frac{i\lambda}{4}
\Big[\Psi^\dagger_\alpha(\vec{\xi})\tau_3
\left(\hspace{-0.5mm}1-\hspace{-0.5mm}\alpha_3\right)\gamma_5
\Psi_\alpha(\xipl) {\rm e}^{-i\omega\lambda}
\nonumber \\* &&\hspace{3.0cm}
-\Psi^\dagger_\alpha(\vec{\xi})\tau_3
\left(\hspace{-0.5mm}1-\hspace{-0.5mm}\alpha_3\right)\gamma_5
\Psi_\alpha(\ximl){\rm e}^{i\omega\lambda}\Big]
\left(\frac{\omega+\epsilon_\alpha}
{\omega^2-\epsilon_\alpha^2+i\epsilon}\right)_{\rm p}
\nonumber \\ &&\hspace{0cm} 
+\sum_\beta\left(\frac{(\omega+\epsilon_\alpha)
(\omega+\epsilon_\beta)}
{(\omega^2-\epsilon_\alpha^2+i\epsilon)
(\omega^2-\epsilon_\beta^2+i\epsilon)}\right)_{\rm p}
\langle \alpha|\tau_3|\beta\rangle
\nonumber \\ &&\hspace{1.0cm} \times
\left[\Psi^\dagger_\alpha(\vec{\xi})
\left(\hspace{-0.5mm}1-\hspace{-0.5mm}\alpha_3\right)\gamma_5
\Psi_\alpha(\xipl) {\rm e}^{-i\omega\lambda}
+\Psi^\dagger_\alpha(\vec{\xi})
\left(\hspace{-0.5mm}1-\hspace{-0.5mm}\alpha_3\right)\gamma_5
\Psi_\alpha(\ximl){\rm e}^{i\omega\lambda}\right]
\Bigg\}
\nonumber 
\ee
and
\be
g_T(x)&=&{\rm eq~(\protect\ref{gtx})} +
\frac{5i}{6\alpha^2}M_NN_C\int \frac{d\omega}{2\pi}
\sum_\alpha \int \frac{d\lambda}{2\pi}\,
{\rm e}^{iM_Nx\lambda}\int d^3 \xi
\label{gttot} \\ &&\hspace{0cm} \times \Bigg\{
\frac{i\lambda}{4}
\Big[\Psi^\dagger_\alpha(\vec{\xi})\tau_1\alpha_1\gamma_5
\Psi_\alpha(\xipl) {\rm e}^{-i\omega\lambda}
\nonumber \\* &&\hspace{3.0cm}
-\Psi^\dagger_\alpha(\vec{\xi})\tau_1\alpha_1\gamma_5
\Psi_\alpha(\ximl){\rm e}^{i\omega\lambda}\Big]
\left(\frac{\omega+\epsilon_\alpha}
{\omega^2-\epsilon_\alpha^2+i\epsilon}\right)_{\rm p}
\nonumber \\ &&\hspace{0cm}
+\sum_\beta\left(\frac{(\omega+\epsilon_\alpha)
(\omega+\epsilon_\beta)}
{(\omega^2-\epsilon_\alpha^2+i\epsilon)
(\omega^2-\epsilon_\beta^2+i\epsilon)}\right)_{\rm p}
\langle \alpha|\tau_3|\beta\rangle
\nonumber \\ &&\hspace{1.0cm} \times
\left[\Psi^\dagger_\alpha(\vec{\xi})\gamma_1
\Psi_\alpha(\xipl) {\rm e}^{-i\omega\lambda}
+\Psi^\dagger_\alpha(\vec{\xi})\gamma_1
\Psi_\alpha(\ximl){\rm e}^{i\omega\lambda}\right]
\Bigg\}\,,
\nonumber
\ee
after taking the collective coordinate (\ref{collq3}) 
and (\ref{collq4}). As it is the case for static properties we 
find that only either the iso--scalar or iso--vector contribution
to any structure function become regularized but not both. For the 
polarized structure function the result that only the 
iso--vector part is regularized was anticipated because we 
know how to relate their zeroth moments to nucleon charges. 
In case of the unpolarized structure function it comes as a 
surprise that the iso--scalar part rather than the iso--vector 
piece undergoes regularization. However for this structure function 
we do not have a normalized sum rule at hand. As discussed, for 
unpolarized structure function which enters the Adler sum rule the
situation is opposite.

\end{document}